\documentclass[aps, prx,twocolumn,longbibliography,superscriptaddress,amsmath,amssymb]{revtex4}
\usepackage[latin9]{inputenc}
\usepackage{amssymb}
\usepackage{graphicx}
\usepackage{amsmath}
\usepackage{color}
\usepackage{mathrsfs}
\usepackage{float}
\usepackage{indentfirst}
\usepackage{mathrsfs}
\usepackage{float}
\usepackage{indentfirst}
\usepackage{textcomp}
\usepackage{comment}
\usepackage{mathtools}
\usepackage{natbib,hyperref}
\usepackage{soul}

\begin{document}

\title{Coexistence of superconductivity with partially filled stripes in the Hubbard model}

\author{Hao Xu}
\thanks{These two authors contributed equally to this work.}
\affiliation{Department of Physics, College of William and Mary, Williamsburg, Virginia 23187, USA}

\author{Chia-Min Chung}
\thanks{These two authors contributed equally to this work.}
\affiliation{Department of Physics, National Sun Yat-sen University, Kaohsiung 80424, Taiwan}
\affiliation{Center for Theoretical and Computational Physics, National Sun Yat-Sen University, Kaohsiung 80424, Taiwan}
\affiliation{Physics Division, National Center for Theoretical Sciences, Taipei 10617, Taiwan}

\author{Mingpu Qin}
\affiliation{Key Laboratory of Artificial Structures and Quantum Control, School of Physics and Astronomy, 
Shanghai Jiao Tong University, Shanghai 200240, China}

\author{ Ulrich Schollw\"{o}ck}
\affiliation{Arnold Sommerfeld Center for Theoretical Physics, Ludwig-Maximilians-Universit\"{a}t M\"{u}nchen, 80333 Munich, Germany}
\affiliation{Munich Center for Quantum Science and Technology (MCQST), 80799 Munich, Germany}

\author{Steven R. White}
\affiliation{Department of Physics and Astronomy, University of California, Irvine, California 92697, USA}

\author{Shiwei Zhang}
\affiliation{Center for Computational Quantum Physics, Flatiron Institute, New York, NY 10010, USA}

\begin{abstract}

Combining the complementary capabilities of two of 
the most powerful modern computational methods, we find 
superconductivity in both the electron-
and hole-doped regimes of the two-dimensional Hubbard model (with next nearest neighbor hopping). 
In the electron-doped regime,  
superconductivity is weaker and is 
accompanied by antiferromagnetic N\'eel correlations at low doping. 
The strong superconductivity on the hole-doped side coexists with 
stripe order, which persists into the overdoped region with weaker hole density modulation. 
These stripe orders, neither filled as
in the pure Hubbard model (no next nearest neighbor hopping) nor 
half-filled as seen in previous state-of-the-art calculations, vary in fillings between 0.6 and 0.8. 
The resolution of the tiny energy scales separating competing orders requires 
exceedingly high accuracy combined with averaging and
extrapolating with a wide range of system sizes and boundary conditions. 
These results validate the applicability of
this iconic model for describing cuprate 
high-$T_c$ superconductivity. 

\end{abstract}

\maketitle

\section{Introduction}

Does the Hubbard model qualitatively capture the essential physics of the high temperature superconducting cuprates? This question has been debated since shortly after these materials were discovered
\cite{Bednorz1986,doi:10.1098/rspa.1963.0204,doi:10.1126/science.235.4793.1196,PhysRevLett.58.2794,PhysRevB.37.3759,doi:10.1063/1.881261,RevModPhys.66.763,2015Natur.518..179K,doi:10.1146/annurev-conmatphys-090921-033948,doi:10.1146/annurev-conmatphys-031620-102024}.
As the decades have passed it has become clearer that the answer has to come from simulations powerful enough to give definitive results on the properties of the model, so that one can see whether these properties match those observed experimentally. This has proved to be especially difficult because the ground states of the models have been shown to be exceptionally sensitive to small changes in the model terms and parameters, with competing \cite{PhysRevX.10.031016} or cooperating \cite{doi:10.1073/pnas.2109406119} charge, spin \cite{PhysRevLett.91.136403}, and superconducting (SC) orders \cite{PhysRevLett.110.216405,PhysRevB.98.205132,PhysRevLett.88.117001,PhysRevB.100.195141,PhysRevLett.113.046402}.  The relevant model parameters are in the 
most difficult regime -- moderately strongly-coupled -- 
where most approaches struggle.  
The frequent presence of stripes in the ground states 
increases the sizes of the clusters needed to extrapolate to the thermodynamic limit. 

A powerful tool has emerged to help overcome these difficulties:  the use of combinations of simulation methods with complementary strengths and weaknesses\cite{PhysRevX.5.041041}.   The density matrix renormalization group (DMRG) \cite{PhysRevLett.69.2863,PhysRevB.48.10345,RevModPhys.77.259} provides the most accurate and reliable results when applied on fairly narrow cylinders \cite{PhysRevResearch.2.033073}.  Other methods work either directly in the thermodynamic limit \cite{PhysRevLett.101.250602,PhysRevLett.81.2514} or at least on much wider clusters \cite{PhysRevB.55.7464}, but have approximations tied to unit cell size\cite{RevModPhys.68.13,PhysRevLett.101.250602,PhysRevLett.109.186404}, coupling strength, etc \cite{PhysRevLett.81.2514,RevModPhys.77.1027,RevModPhys.84.299}. 
The constrained path (CP) auxiliary field quantum Monte Carlo (AFQMC) method \cite{PhysRevB.55.7464,PhysRevB.78.165101,PhysRevB.94.235119} is particularly complementary to DMRG: it can be used on much wider systems; 
the errors from CP to control the sign problem have been consistently modest  \cite{PhysRevX.5.041041};  
and the underlying approximation of CP is
unrelated to the low entanglement approximation of DMRG. 
AFQMC is based on a wave picture of superposition of Slater determinants, while
DMRG is rooted in the particle picture with strong coupling. 
Their quantitative handshake 
proved to be crucial for uncovering the delicate nature of the stripe correlations as we discuss below.  
Previously, we used this combination, extrapolating to the two-dimensional thermodynamic limit, to find that superconductivity is absent in the pure 
(i.e., with no next nearest-neighbor hopping)
Hubbard model \cite{PhysRevX.10.031016}. In that case, the lack of superconductivity was tied to the occurrence of filled striped states\cite{Zheng1155}.

Here, we 
apply this approach, with new developments, to tackle
the Hubbard model with 
a non-zero next nearest-neighbor hopping, $t'$.
In connection to the typical phase diagram of cuprates, a nonzero $t'$ is necessary to account for the particle-hole asymmetry and the band structures.
The $t' \ne 0$ model is significantly more difficult computationally, with challenges for both DMRG and AFQMC. 
Where both methods apply, DMRG certifies the high accuracy and reliability of AFQMC as used here. 
As discussed below, in cases of ambiguity (e.g., in some width-6 cylinders), resolving the discrepancies has often created new synergy between the two methods, and led to new insights. The phase diagram with $t'$ also turns out to be significantly more complicated,
with partially filled stripes coexisting with superconductivity on the hole-doped side, and uniform antiferromagnetic order coexisting with superconductivity on the electron side. 
The final results for superconductivity, extrapolated to the thermodynamic limit, are impressively similar to the properties of cuprates, with both electron and holed doped SC ``domes", but with the hole doped side being significantly stronger.

The Hamiltonian of the Hubbard model is  
   \begin{equation}  \label{eq:H}  
   \begin{split}
   \hat{H} =-t\sum\limits_{\langle ij\rangle,\,\sigma}\hat{c}_{i\sigma}^{\dagger}\hat{c}_{j\sigma}  
 -t' \sum\limits_{\langle\langle ij\rangle\rangle,\,\sigma}  \hat{c}_{i\sigma}^{\dagger}\hat{c}_{j\sigma}    \\
     +U\sum\limits _{i}\hat{n}_{i\uparrow}\hat{n}_{i\downarrow}
     -\mu\sum_{i\sigma}\hat{n}_{i\sigma}  
     \end{split}
     \end{equation}  
where $i$ or $j$ labels a site on a square lattice, 
$\hat{c}^\dagger_{i\sigma}$ is the electron creation operator, $\sigma=\{\uparrow,\downarrow\}$ denotes spin, $\hat{n}_{i\sigma}=\hat{c}_{i\sigma}^\dagger\hat{c}_{i\sigma}$ is the particle-number operator,
and $\langle ij\rangle$ and $\langle\langle ij\rangle\rangle$ indicate 
nearest- and next-nearest-neighbors, respectively.
We set $t$ as the energy unit.
In cuprates $t'<0$ \cite{RevModPhys.75.473}; 
however,using a particle-hole transformation to map fillings $1+\delta \to 1-\delta$, 
we can study electron doping by changing
the sign of $t'$.
We use $t' = -0.2$ for hole-doping 
and  $t' = + 0.2$ for electron-doping,
appropriate values
for cuprates based on band structure calculations  \cite{ANDERSEN19951573,PhysRevB.98.134501}. 
The onsite repulsion $U$ is fixed at $U=8$,
again a representative value for cuprates. We scan a range of doping
(denoted by $\delta$) by varying $\mu$.

\begin{figure}[]
  \includegraphics[width=0.98\linewidth]{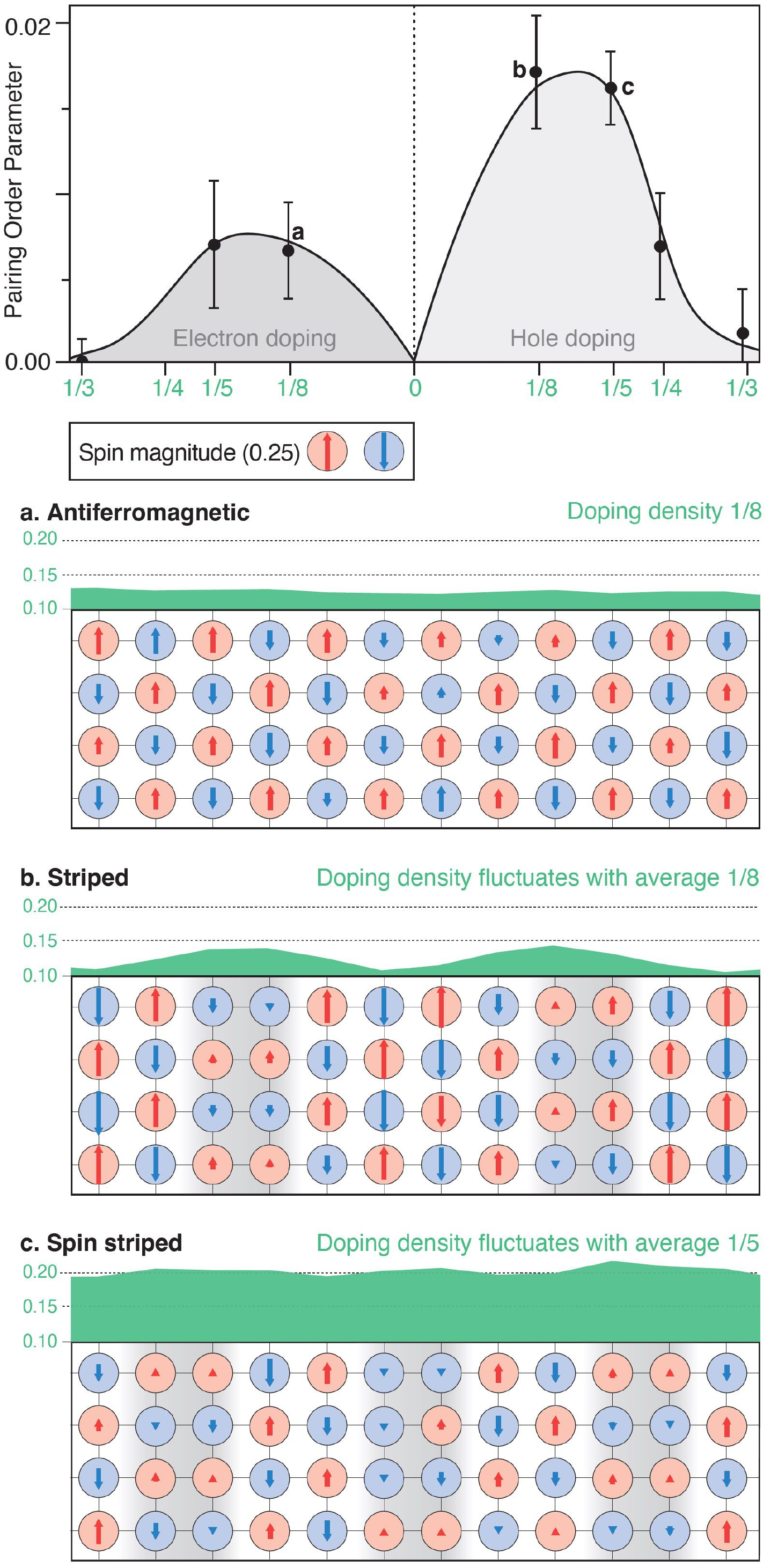} 
  \caption{The $d$-wave pairing order parameter versus doping $\delta$ in the ground state for the hole-doped ($t'=-0.2$) and electron-doped ($t'=+0.2$) regimes.  Representative spin and charge correlations are also shown for three parameter sets a, b, and c. 
$\Delta_d$ are the spontaneous pairing order in the thermodynamic limit, 
while the spin and charge (hole) patterns are drawn from 
the middle of $28\times 8$ (a), $24\times 8$ (b), and $40\times 8$ (c) cylinders 
with antiferromagnetic spin pinning fields applied to the two edges.
Note that hole densities start at $0.1$.  Grey shadows for spins are to aid the eye.}
\label{phase}
\end{figure}

Our study focuses on the ground state, which we 
obtain in either cylindrical or fully periodic systems. 
The use of cylinders serves two purposes. First they allow direct comparisons between AFQMC and DMRG, which is highly accurate in narrow cylinders. Second, they are convenient for studying spin and charge orders, in which we apply spin-symmetry-breaking pinning fields on the edges of the cylinder to help detect ordering from the resulting local spin and charge densities. 
The fully periodic simulation cells 
allow AFQMC to better approach the thermodynamic limit (TDL).  As shown below, it turns out to be crucial to 
systematically average over different boundary conditions.  
To compute the pairing order parameter, we apply twist averaged boundary conditions (TABC) over a large number of random twists, in systems 
with up to 500  lattice sites.
The computations presented in this work became possible only with new algorithmic developments in both our methods, which improved capability and increased accuracy, as we discuss further in the Method Section. 

\section{Results}

\subsection{Overview of pairing and coexisting spin/charge orders}

Figure \ref{phase} presents an overview of our results, a ``phase diagram'' of the computed pairing order parameter, together with 
representative spin and charge correlations.
The pairing order parameters have been extrapolated to the TDL, 
using full TABC in large simulation cells 
(see Method and SM).
We expect this zero-temperature property to be loosely connected
to the transition temperature $T_c$ most readily observed experimentally (however,
see \cite{foot2,1995Natur.374..434E}). 
On both the electron- and hole-doped sides, we find dome-like $d$-wave pairing orders which resemble the $T_c$ domes in the typical phase diagram of cuprates. 
The pairing order is significantly larger in the hole-doped region than in the electron-doped region, which is also consistent with the phase diagram of cuprates \cite{RevModPhys.84.1383}. 
Spin and hole densities are shown for the three representative systems marked as a, b, and c.
These calculations were performed with AFM pinning fields on the edges of the cylindrical simulation cells
(details in SM).
The spin and 
hole densities thus provide a simple and convenient way to visualize the spin and charge correlations. 
We have taken care to 
ensure that the results are drawn from very large systems 
and the spin and charge patterns
are representative of different boundary conditions.
In the electron-doped region, the spins show single-domain antiferromagnetism  with nearly uniform hole densities in the bulk. 
In the hole-doped region, stripe and spin-density wave (SDW) correlations are observed, with modulated antiferromagnetic 
domains separated by phase flip lines where holes are more concentrated. 
In contrast with the pure Hubbard model, we find that the wavelength 
of the modulation is not an integer multiple of $1/\delta$ (filled stripes). Nor are the stripes half-filled as seen in previous state-of-the-art calculations
\cite{Huang_2018}. 
Rather, they are best described as partially filled,
with fractional fillings which vary with $\delta$ as well as system size and boundary conditions.
These behaviors of spin and charge 
are again consistent with the phase diagram of the cuprates \cite{RevModPhys.84.1383}, where uniform AF correlations persist with substantial doping on the electron-doped side, but {short or long-ranged} incommensurate magnetism and stripes are observed starting at small doping on the hole-doped side \cite{nature_375_15_1995,doi:10.1080/00018732.2021.1935698}. 

This phase diagram contrasts sharply with 
that of the $t$-$t'$-$J$ model\cite{doi:10.1073/pnas.2109978118,PhysRevLett.127.097003}, which can be derived as an approximate strong-coupling Hubbard model at low doping. 
In the $t$-$t'$-$J$ model, recent 
DMRG studies all point to strong $d$-wave superconductivity on the electron-doped side 
\cite{doi:10.1073/pnas.2109978118,PhysRevLett.127.097003,PhysRevLett.127.097002}, which coexists with antiferromagnetic correlations with increasing strength as $t'$ increases; some differences remain concerning whether long-range AF order occurs \cite{Kivelson-t-tp-J-preprint}.
No superconductivity, only stripes, have been found on the hole-doped side. 
It has been an open question whether this failure of the  $t$-$t'$-$J$ model to qualitatively explain the cuprates was due to the strong-coupling approximations of that model, or to other flaws or missing terms affecting both the Hubbard and $t$-$t'$-$J$ (single band) models. Here the strong differences in the phase diagrams of the two models point to the former.
These differences have not been clear in previous studies on narrower cylinders, 
which are impacted by strong finite-size effects
\cite{2019Sci...365.1424J,PhysRevB.102.041106}.

\subsection{Underdoped region: $1/8$ hole doping}
\begin{figure}[t]
        \centering{
	\includegraphics[width=0.98\linewidth]{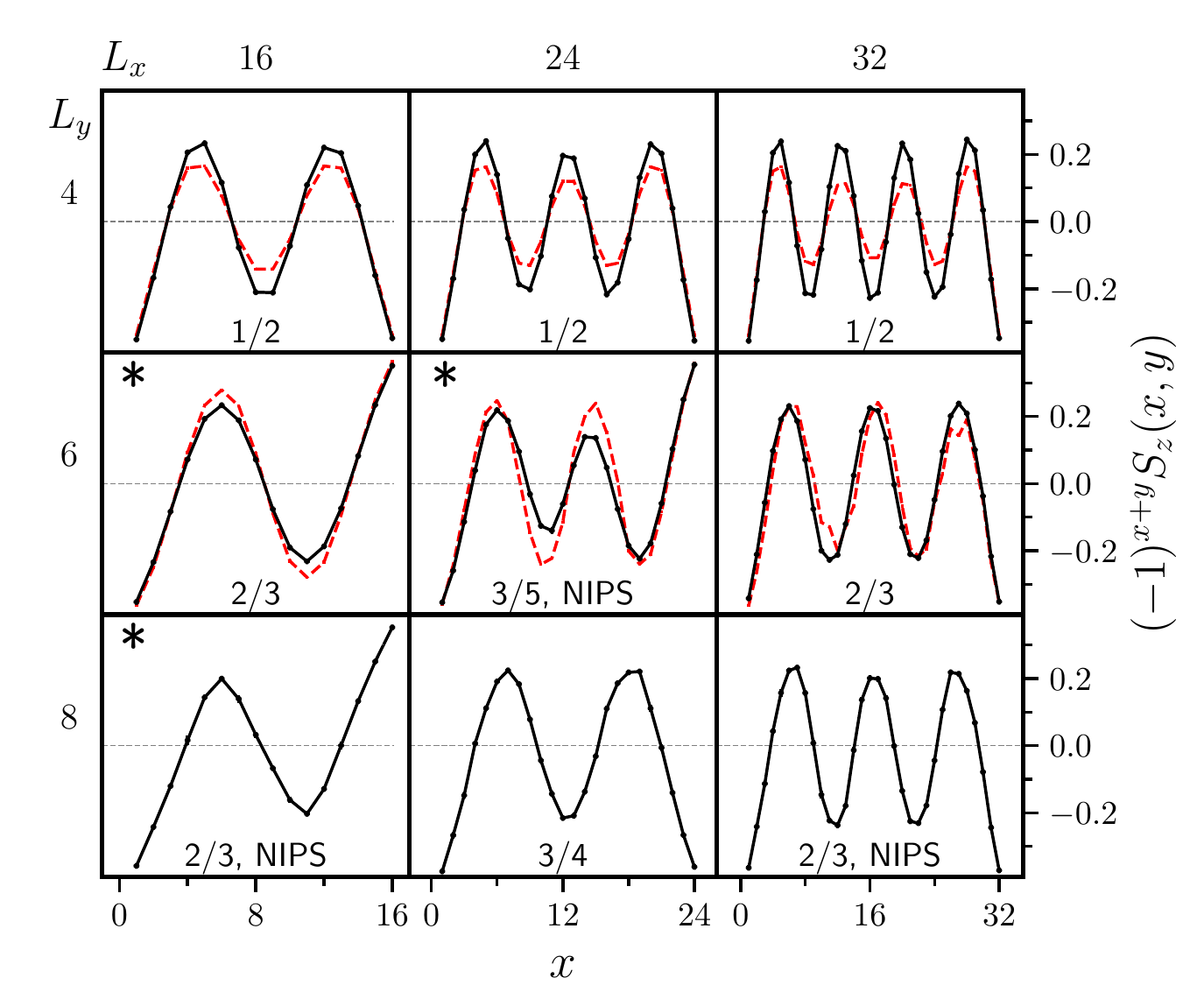}
        }
	\caption{
 Evolution of the stripe patterns with system size ($\delta=1/8$, hole-doped).
 The staggered spin densities are shown as linecuts in periodic cylinders. The length of the cylinder ($L_x$) is varied across 
 the three columns  
and the width ($L_y$) across rows. 
AFM pinning fields are applied at the two edges of the cylinder ($x=1$ and $x=L_x$), either in phase or with a $\pi$-phase shift
(marked by an asterisk);
the one with lower energy is shown.
The filling fraction $f$ of each stripe pattern is indicated, with NIPS denoting non integer-pair stripes. 
 DMRG results (red) are shown for width-4 and 6 systems and AFQMC results (black) are in good agreement with them.
}
\label{stripe-hole}
\end{figure} 

A relatively large pairing order parameter is found 
here, in coexistence with stripe correlations, 
as shown in Fig.~\ref{phase}.
To better understand the nature of the spin and 
charge correlations, we systematically study
their evolution with system sizes in Fig.~\ref{stripe-hole}.
The computations 
were performed in $L_x\times L_y$ 
cells, with periodic (PBC) 
or anti-periodic boundary condition (APBC)  in the $\hat y$-direction and open BC along 
$\hat x$ (i.e., cylinders). AFM pinning fields 
(along $\hat z$) were applied at $x=1$ and $L_x$
to break the SU(2) symmetry and induce local spin orders, 
such that the local spin density $S_z(x,y)$ becomes a proxy of spin-spin correlations away from the edges of the cylinder.  

Modulated AFM patterns are clearly seen in all the systems. 
Correspondingly, hole densities are enhanced at the nodes of the spin modulation, 
as illustrated in Fig.~\ref{phase} (results on the corresponding hole densities for Fig.~\ref{stripe-hole} can be found in SM).  The characteristic wavelength 
of the modulation, $\lambda_{\rm SDW}$, varies with system size. We define a filling fraction of the 
stripe: $f\equiv \delta\, \lambda_{\rm SDW}/2$, i.e., the number of holes per lattice spacing along a stripe. In the pure Hubbard model, $f=1$ since $\lambda_{\rm SDW}=2/\delta$ \cite{PhysRevLett.104.116402,PhysRevResearch.4.013239}. Then, \emph{nominally} the number of electron pairs per stripe 
is $n_p\equiv f\,L_y/2$. If $n_p$ is an integer,
we refer to the state as integer-pair stripe (IPS); otherwise the state is labeled as non-IPS (NIPS). 

Previous studies in width-4 cylinders have found that the ground state in this system has half-filled stripes  \cite{Huang_2018,doi:10.1073/pnas.2109978118,PhysRevLett.127.097003}. 
Our results confirm this picture, with good agreement between AFQMC and DMRG,
but also show that the half-filled stripe turns out to be special to width-4. 
As the system size increases, 
the 
stripe filling fluctuates between $3/5$ and  $3/4$. NIPS states 
appear frequently, which have not been observed before. 
Previous calculations \cite{PhysRevB.60.R753,PhysRevX.10.031016} show that states with IPS
are favored, which was taken as an indication of the existence of local pairing of electrons in the stripe state. Here, with the inclusion of $t'$, the electron is more mobile and pairs of electrons become coherent to display long-range pairing order. 
This is further discussed and contrasted with 
the over-doped region next.

\subsection{Overdoped region: $1/5$ hole doping}

\begin{figure}[t]
        \centering
        {

    \includegraphics[width=0.98\linewidth]{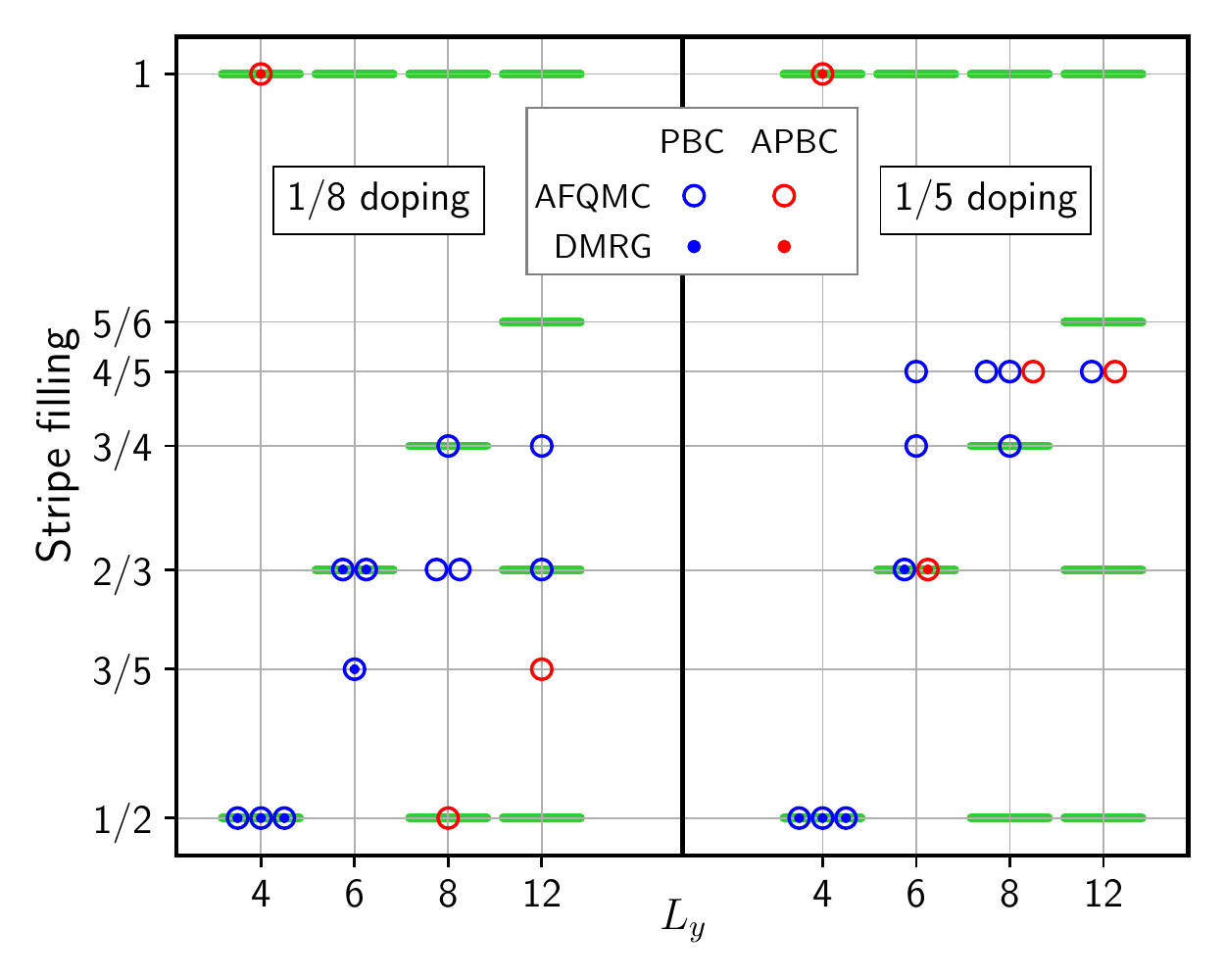}
	}
         \caption{Partially filled stripe patterns on the hole-doped side,
         at $\delta= 1/8$ and $1/5$. 
         The stripe fillings are shown for a variety of 
         system sizes, in cylindrical cells with width $L_y=4$ up to $12$, and lengths 
         ranging from $16$ to $48$
         (shown as adjacent symbols at fixed $L_y$).  
         Results for both PBC and APBC are shown.
         Narrow cylinders favor integer-pair stripes (IPS, indicated by green bars). Fluctuations
         are strong even in large systems. 
         } 
	\label{stripe-hole-over}
\end{figure} 

A strong superconducting order parameter is found in the ground state 
of the hole overdoped region of $\delta=1/5$, 
with strength comparable to $\delta=1/8$ (see Fig.~\ref{phase}). 
The behavior of spin and charge correlations show common features but also significant differences between the two regions. 
Figure~\ref{stripe-hole-over} summarizes 
their stripe fillings side by side, based on 
computations in about $30$ systems.
Several trends are evident. In narrow cylinders, IPS states are favored at both dopings. 
In over a dozen different width-4 and width-6 systems across 
the two dopings, 
AFQMC and DMRG agree in each case on the stripe 
wavelength and filling fraction. 
In both regimes
the filling fraction varies widely with
system sizes and boundary conditions, and fluctuations continue through systems 
with over $500$ lattice sites.  
As the size grows (wider cylinders), IPS states are no longer favored, and both systems tend to fractional stripe fillings. 
These results indicate that with $t'$, the stripe patterns --- but not the existence of stripes --- are much more fragile than in the pure Hubbard model.  

Both the spin and charge modulations are weaker at $1/5$ doping than 
at $1/8$. 
Although $f$ is larger in the TDL,
the holes are more mobile and spread out in the overdoped region. 
The hole density is nearly uniform, with less than $5$\% of the holes
contributing to the density fluctuations.
At $1/8$ doping, the stripe order is more 
pronounced, as illustrated in Fig.~\ref{phase}. 
Still, the peak density of holes, at the nodes of the spin correlation, is 
only $\sim 30$\% higher than the average. 
The notion of stripe filling derives from a particle picture, most applicable 
to holes in Wigner-crystal-like distributions. 
The holes here have 
a strong wave character \cite{PhysRevLett.104.116402}, with which the fractional fillings of stripes we observe are more readily compatible.

\subsection{Electron doped region}

\begin{figure*}[t]{

	\includegraphics[width=0.9\linewidth]{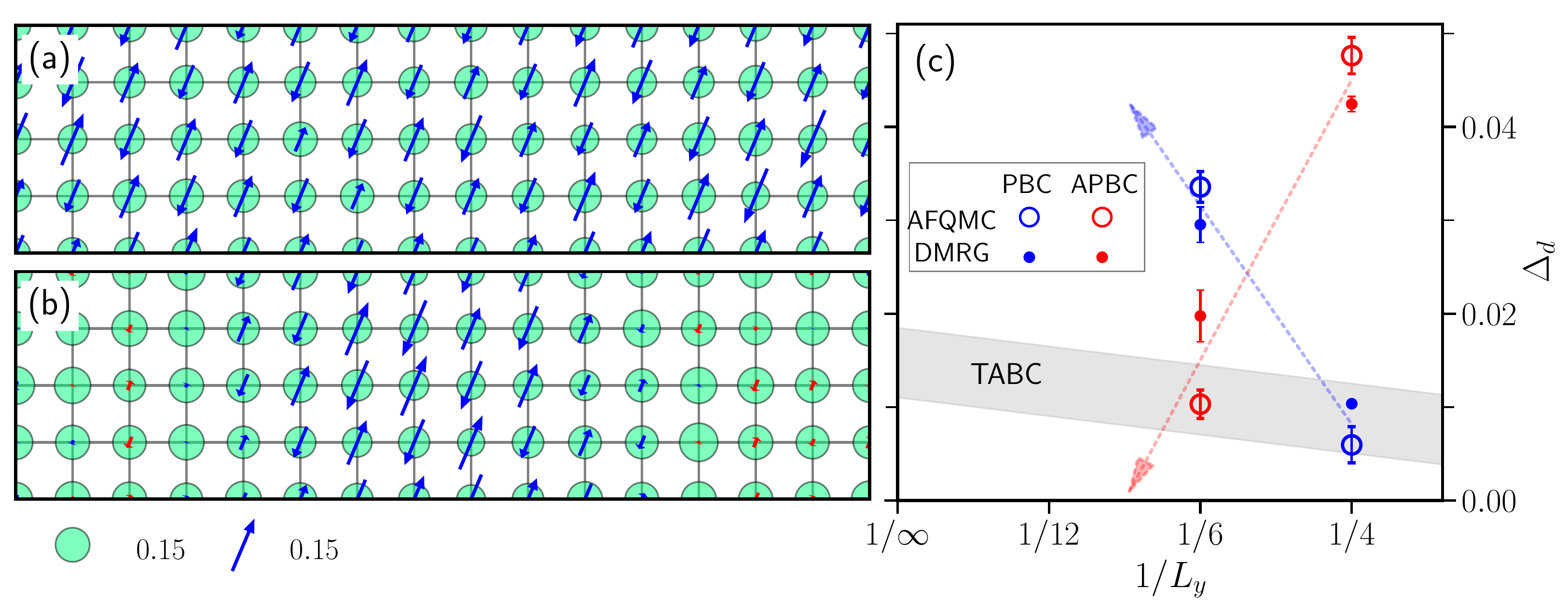}
	
         \caption{Spin, charge, and pairing 
         properties on the electron doped side ($\delta= 1/8$),
          and their variations with 
          boundary conditions.  (a) APBC along $ \hat y$-direction in a $28\times 8$ cylinder  gives 
          nearly uniform Neel order (only a $16\times4$ central region is shown). 
         (b) Under PBC 
         a modulated AFM order with larger spatial variations in spin magnitude is seen. 
        (c) The computed pairing orders in
        $16\times 4$ and $16\times 6$ cylinders
        (at a fixed value $h_d=0.021$ 
        of applied global $d$-wave pairing fields) 
        show opposite trends with PBC and APBC.
        The final pairing order, computed from  
        TABC with fully periodic supercells of increasing $L_y$, is shown together with the TDL extrapolation by the gray band.
    }  
	\label{twist-pos}

}
\end{figure*}

Experimentally, the electron-doped side is simpler,
without the competing stripe state \cite{nature_375_15_1995,RevModPhys.87.457} or pseudogap phase in cuprates \cite{RevModPhys.84.1383}. The critical doping for the long-range AF order on the electron-doped side 
is larger than that on the hole-doped side,  the superconducting dome 
is smaller, and the transition temperature is lower. The phase diagram in Fig.~\ref{phase} and the spin and hole densities in Fig.~\ref{twist-pos} are consistent with these features.

Our results reveal several other important features on the electron-doped side. 
There are considerable variations 
of the spin and charge correlations
with system sizes and boundary conditions,
even though the sensitivity is less compared to the
hole-doped side.
As illustrated in the SM, 
two entirely different ground-state orders are obtained from width-4 and width-6 cylinders; 
APBC and PBC also 
lead to opposite conclusions in each simulation cell.
Even in the 
 width-8 systems
 in Fig.~\ref{twist-pos}, which display robust 
 N\'eel order, different boundary 
conditions 
still show variations in the
charge correlation. Superconductivity manifests a more dramatic volatility. 
Using PBC, the most common approach to date,
calculations in width-4 and width-6 
cylinders 
would conclude a strong 
pairing order in the electron-doped regime.
(Note that DMRG and AFQMC give fully consistent results.)
In contrast, under APBC
the same calculations predict no pairing. 
The uncertainties with respect to finite size and boundary conditions
are much larger than the final signal at the TDL.
Thus even a qualitative conclusion on superconductivity would be challenging 
without our new approaches employing
TABC, systematic extrapolation to large sizes, and other methodological advances, which are discussed next.

\section{Method}
\label{method}

The physics of the Hubbard model has proved highly elusive and challenging to pin down. This was magnified substantially with a non-zero $t'$. 
The difficulties include more 
sensitivity and stronger dependency on system size and BC, as we have illustrated. In addition, $t'$ turns out to affect the interplay between low-lying states  in significant ways. 
For instance, with $t'=0$, stripe and superconductivity manifest as competing orders. Filled stripe states are particularly stable, with nesting contributing a key factor.
A non-zero $t'$ affects the nesting condition (frustrates the  N\'eel order) and alters the landscape of the low-lying states. This has demanded much higher resolution from the numerical methods. 

The methodologies employed in this work have a number of distinguishing 
features which made it possible to achieve a qualitatively higher level of accuracy and reliability. 
Two complementary, state-of-the-art computational methods are used synergistically. 
We implement both U(1) \cite{10.21468/SciPostPhysCodeb.4} and SU(2) symmetry-adapted \cite{hubig:_syten_toolk} DMRG calculations for different setups and push them to the large bond-dimension limit. 
In AFQMC, we introduce a further advance in the optimization of the constraining trial wave function,  
which is determined fully self-consistently \cite{PhysRevB.94.235119}, 
with no input parameter.
Extensive and detailed comparisons between AFQMC and DMRG are performed on width-4 and width-6 cylinders, under identical conditions. The same AFQMC algorithm, which has no room for tuning, is applied to larger systems.
The formulation of systematic twist averaging for the 
computation of the pairing order parameters
provides an effective way to sample the low-lying states.

\subsection{Twist averaging as an effective means to sample low-lying states}

\begin{figure}[t]
        \centering{
	\includegraphics[width=0.98\linewidth]{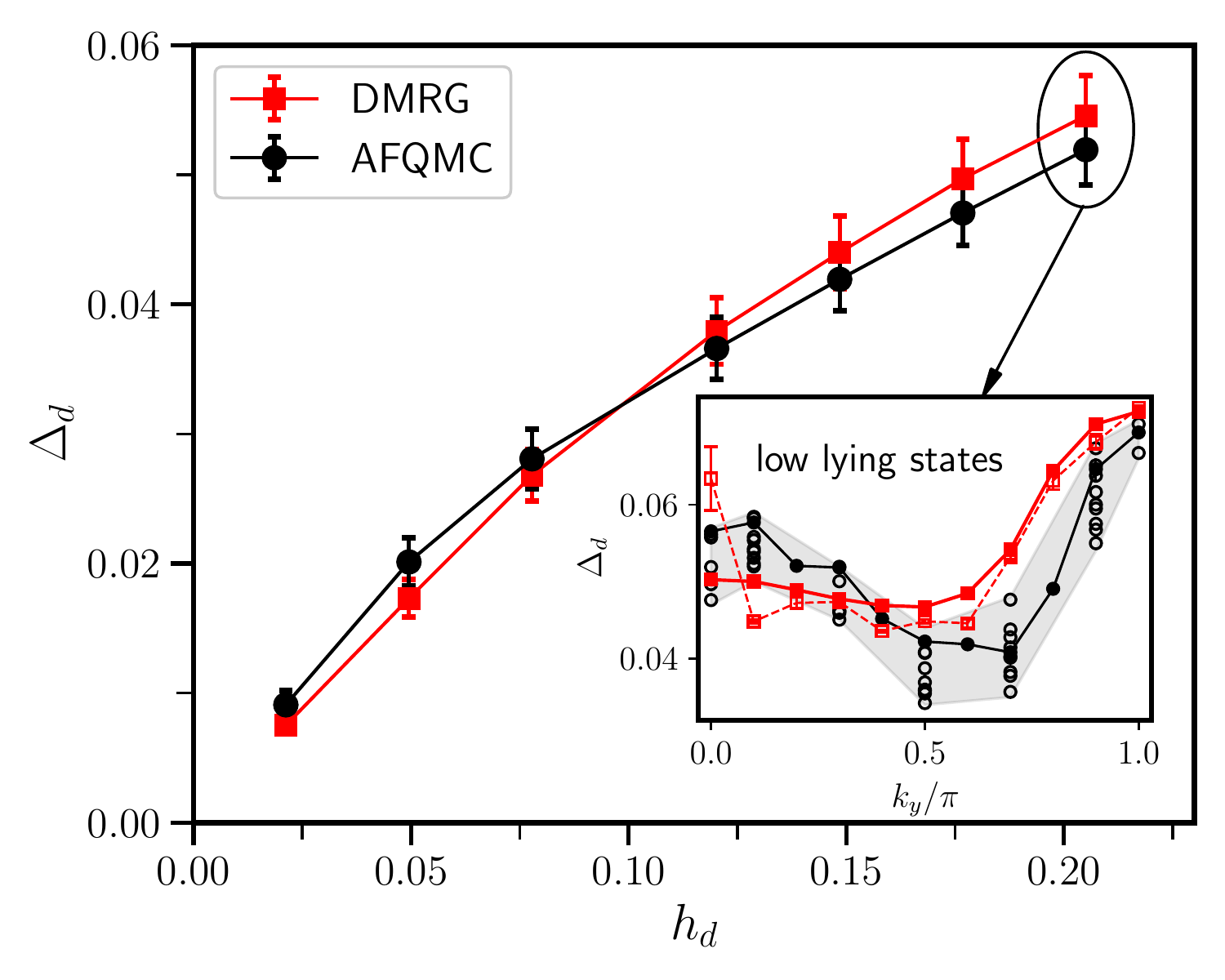}
	}
         \caption{Importance of TABC for accurate determination of the pairing order. 
         The main figure shows the $d$-wave
         pairing order parameters 
         in a $20 \times 4$ cylindrical cell at $1/5$ hole doping, 
          after full twist-averaging over $k_y$. 
         AFQMC and DMRG results  
          agree across the entire range of $h_d$, 
          the strength of the applied pairing fields.
          The inset focuses on $h_d=0.205$. $\Delta_d$ computed from DMRG and AFQMC are shown as a function of $k_y$, for the ground state 
          (connected by solid line)
          and some of the lowest-lying excited states (open symbols).  Averages of the solid symbols lead to the TABC results in the main figure. 
       }
	\label{twist}
\end{figure}

The use of twist-averaging \cite{PhysRevE.64.016702,PhysRevB.94.085103} in this work has two crucial roles. First, systematically averaging over twist angles,  combined with the ability to reach large 
system sizes 
and careful finite size extrapolation,
enables us to approach the TDL reliably. 
 Second, the random twist angles provide an effective means to sample the low-lying states, and their averaging reduces the impact of
rare events of accidental degeneracy, and smoothes out the effect of level crossings as a function of an applied pairing field (see SM).

As shown in Fig.~\ref{twist-pos}, different boundary conditions can result in variations in the pairing order parameter
which are many times larger than the signal,
even in nominally rather large sizes 
(width-6 cylinders).
Both PBC and APBC are twist angles of special symmetry, and are often particularly volatile. 
We apply TABC with quasi-random twist angles 
\cite{PhysRevB.94.085103}. 
The TBC can be thought of as 
the electron gaining a phase when it crosses the boundary.
Equivalently, we can choose another gauge by distributing the phase evenly in each hopping term.
When a twist is applied, care must be taken in defining the pairing order parameter, whose 
form is gauge-dependent but the expectation value should be gauge-independent. 
TABC reduces the fluctuations in the computed pairing order parameter, as seen in
Fig.~\ref{twist-pos}, and further discussed below and in the SM. 
(In Ref.~\cite{PhysRevB.107.075127}, TBC and twist averaging are shown to accelerate the extrapolation with calculations on cylinders.)

With the inclusion of 
a non-zero $t'$, the perfect nesting in the Fermi surface at half-filling is absent. Subtle variations 
near the Fermi level from finite size and boundary conditions can have much larger effect on
the formation of collective spin modes,
hence there is more sensitivity in the property 
of the low-lying states. 
These states can be very close  
in energy such that any small finite 
temperature (e.g., under experimental conditions) would smear them out and render them indistinguishable. TABC provides 
an effective sampling of such low-lying states which can average  
out the
fluctuations 
so as to more reliably
capture the intrinsic properties.  
An illustration is given in Fig.~\ref{twist}.
The pairing order parameter exhibits large variations 
as a function of the twist angle, 
both in the ground state and low-lying excited states, as seen in the inset for one value of $h_d$.
The calculation can ``hop'' from one state to another among the bundle of low-lying states,
depending on the initial condition, convergence 
criterion, etc,
even under 
high-quality computational settings (e.g., large 
bond dimensions in DMRG).
This is also reflected in the 
modest level of agreement between the two methods for each 
particular state. With TABC, however, 
their agreement is excellent
across the entire range of $h_d$ (which spans many level-crossings, see SM), and the two methods
give fully consistent conclusions.

\subsection{Extrapolation of pairing order}
 
\begin{figure}[t]

        \includegraphics[width=0.98\linewidth]{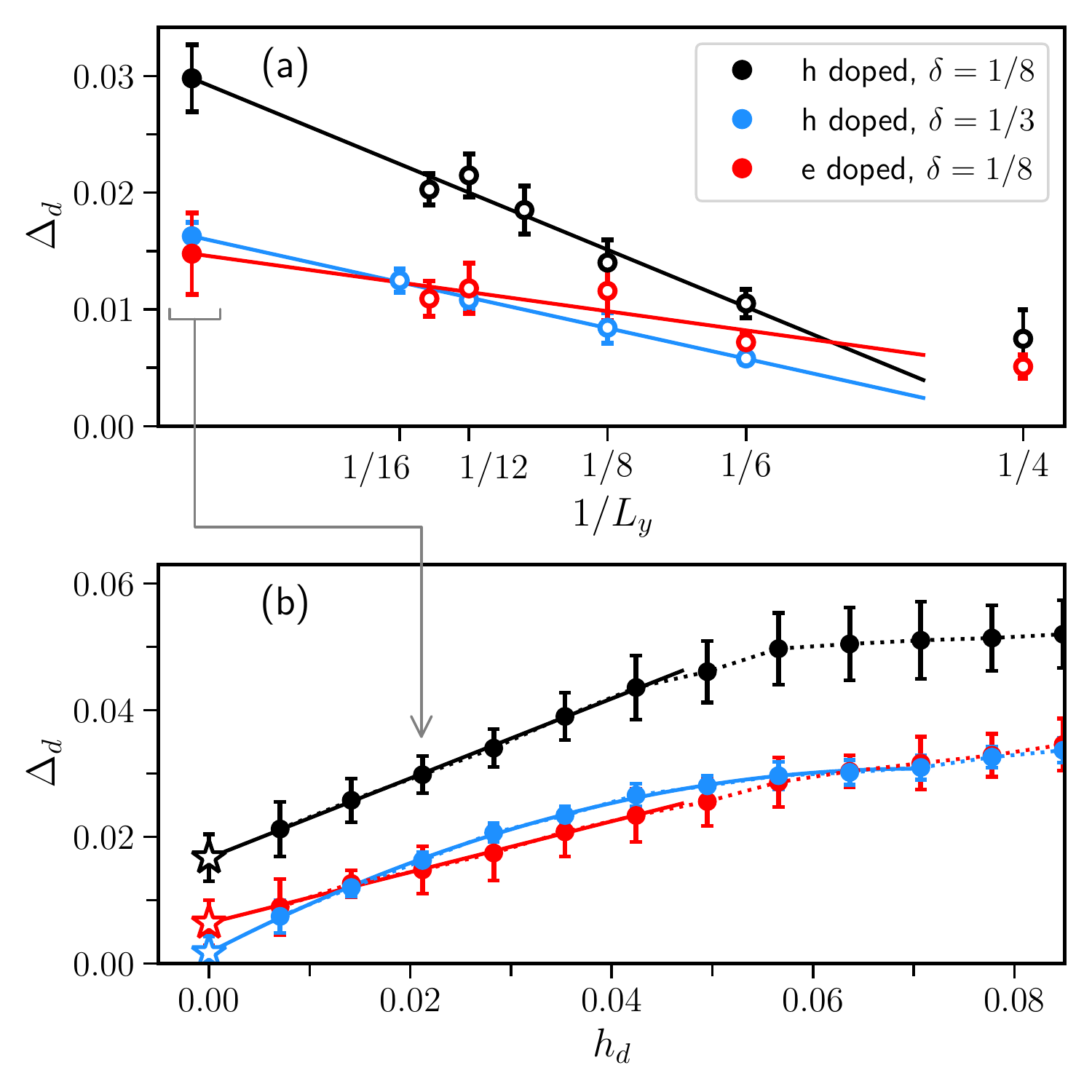}
        \caption{
         Computation of the ground-state pairing order parameter at the thermodynamic limit. 
         (a) shows extrapolation to the TDL at a fixed $h_d$, the strength of the $d$-wave pairing fields. (b) shows extrapolation of the TDL 
         result from (a) to $h_d\rightarrow 0$.
         Three representative systems are shown.
         In (a), each data point is obtained by TABC over $(k_x,k_y)$ in 
         supercells of $L_x\times L_y$, and only results from large supercells 
         are included.  
         In (b) linear or quadratic fits are performed at 
         small values of $h_d$, 
         with extrapolated values marked as stars. 
        }
     
	\label{pair-extra}
\end{figure} 

The spontaneous pairing order parameter in the TDL, $\Delta_d$, is obtained from a massive number of 
computations. At each parameter set ($t'$ and 
doping),
$\Delta_d(N, h_d)$ 
is computed 
for many different 
simulation cell 
sizes $N$, at tens of $h_d$ values, with each averaged over 
tens of quasi-random twist angles. We then 
take the limit $\Delta_d(N\rightarrow \infty, h_d)$ at each 
$h_d$, followed by the extrapolation  
$\Delta_d(\infty, h_d\rightarrow 0)$. The procedure is illustrated in  Fig.~\ref{pair-extra}.
Panel (a) shows the first step, where we use 
fully periodic 
$N=L_x\times L_y$ systems
with quasi-random twist angles $(k_x,k_y)$ applied 
to both directions. We verify that $L_x$ is sufficiently large such that the results have converged within our statistical accuracy. We then extrapolate the TABC results 
 with respect to $1/L_y$, excluding 
small sizes. (Deviations are visible
from width-4 systems,  
which can have different pairing 
symmetry from ordinary $d$-wave \cite{PhysRevB.102.041106}.)
In Panel (b) extrapolations are then performed 
using small $h_d$ values ($<0.05$ for linear 
and last $10$ or so points for quadratic fits),
yielding the final spontaneous pairing order parameter $\Delta_d$ at $h_d\rightarrow 0$. 
As can be seen, the quality of the fits is excellent;
in each case, the linear and quadratic fits give 
consistent values within statistical errors.

\section{Conclusion}
Can the single band Hubbard model capture the qualitative physics, particularly the superconductivity, of the cuprates? Here, more than 35 years after the discovery of the first cuprate superconductor \cite{Bednorz1986},  we conclude that the answer is yes, that the Hubbard model with a next near-neighbor hopping $t'$ distinguishing between electron- and hole-doping captures the essential features of the charge, magnetic, and pairing orders.

The computed pairing order parameter in the ground state
displays dome-like structures 
versus doping,  resembling the $T_c$ domes of the cuprates. On the hole-doped side, we find the coexistence of superconductivity with fractionally filled stripe correlations, with nominal stripe fillings in the range 0.6-0.8 in sufficiently large sizes. 
On the electron-doped side, at lower dopings, uniform or weakly modulated antiferromagnetism, along with uniform or weakly modulated doping, coexists with somewhat weaker superconductivity.  The general appearance of stripe orders on the larger systems with non-integral numbers of pairs indicates that pairs fluctuate between stripes, promoting
long-distance phase coherence and thus superconductivity; in contrast, for $t'=0$ the stripes were
filled, and superconductivity was absent \cite{PhysRevX.10.031016}. 

This picture is in contrast to that of the $t$-$t'$-$J$ model, once thought to be interchangeable with the Hubbard model, but which does not appear to exhibit superconductivity on the hole-doped side \cite{doi:10.1073/pnas.2109978118,PhysRevLett.127.097002,PhysRevLett.127.097003}.
The ground states of the models are not universal, and to capture the subtle interaction of the various 
intertwined orders 
requires both very careful finite size extrapolation and very high accuracy and reliability in the simulation methods. 
Even within the single-band 
$t$-$t'$ Hubbard model, an enormous body of works exists, with widely varying and often conflicting results.
Our results also explain why this has been the case ---
the model shows extreme sensitivity of the properties to finite sizes and boundary conditions, and to any  biases of approximate methods.

Here we have used the combination of  DMRG and AFQMC, with DMRG benchmarking and validating the CP approximation in AFQMC on narrower systems and the AFQMC used 
to reach much larger
systems. We have greatly improved the finite size extrapolations by using TABC. 
These together with methodological advances within 
each approach provided a powerful tool 
to address the question with a new
level of capability and resolution.

In the models or parameter regimes on the hole-doped side where  
superconductivity is not present, one still finds strong indications of paired holes.  
For example, if holes within stripes were not paired, one would expect to find single stripes having an odd number of holes in about half the systems, but 
instead 
only even numbers of holes in each stripe are found.
Whether there is superconductivity or not seems tied to the properties of a pair, 
e.g., its effective mass, which is strongly influenced by model parameters such as $t'$.  A heavy pair or one which interacts strongly with the magnetic degrees of freedom of the region around it is more likely to be locked up in a stripe, suppressing phase coherence. This model-specificity and non-universality raises the question:  is there any simple analytic theory of cuprate superconductivity in the style of BCS, or must we always resort to simulation?

Our study still leaves much to do in connecting the models quantitatively to experiments.  We have not predicted transition temperatures, only order parameters.  We have not studied transport and dynamical properties of the models. 
Many other properties of the one-band Hubbard
model remain to be determined and understood. 
Other terms \cite{doi:10.1126/science.abf5174,DMRG-downfolding} and effects not present in the Hubbard model may still play important quantitative roles.  
Nevertheless, it appears that qualitatively, the $t$-$t'$-$U$ Hubbard model  has ``the right stuff''.

\section{Acknowledgments}
We thank A.~Georges, S.~Kivelson, A.~J.~Millis, M.~Morales, H.~Shi, E.~Vitali, and T.~Xiang for discussions.
We are grateful to Lucy Reading-Ikkanda for help with graphics.
M.Q acknowledges the support from the National Key Research and Development Program of MOST of China (2022YFA1405400), the National Natural
 Science Foundation of China (Grant No. 12274290) and the sponsorship from Yangyang Development Fund. SRW acknowledges the support of the NSF
 through under DMR-2110041. US acknowledges funding by the Deutsche Forschungsgemeinschaft 
 (DFG, German Research Foundation) under Germany's Excellence Strategy-EXC-2111-390814868.
H.X.~thanks the Center for Computational Quantum Physics, Flatiron Institute for support and hospitality. The Flatiron Institute is a division of the Simons Foundation. C.-M.C. acknowledges the support by
the Ministry of Science and Technology (MOST) under Grant
No. 111-2112-M-110-006-MY3, and by the Yushan Young
Scholar Program under the Ministry of Education (MOE) in Taiwan.

\bibliography{main}

\begin{thebibliography}{59}
\bibitem[{foo()}]{foot2}
\bibinfo{note}{A better proxy for $T_c$ might be the magnitude of the
  superconducting gap. This would be more difficult to calculate, and it also
  is only roughly tied to $T_c$.}
\end{thebibliography}

\end{document}


\title{Supplementary materials for ``coexistence of superconductivity with partially filled stripes in the Hubbard model"}

\author{Hao Xu}
\thanks{These two authors contributed equally to this work.}
\affiliation{Department of Physics, College of William and Mary, Williamsburg, Virginia 23187, USA}

\author{Chia-Min Chung}
\thanks{These two authors contributed equally to this work.}
\affiliation{Department of Physics, National Sun Yat-sen University, Kaohsiung 80424, Taiwan}
\affiliation{Center for Theoretical and Computational Physics, National Sun Yat-Sen University, Kaohsiung 80424, Taiwan}
\affiliation{Physics Division, National Center for Theoretical Sciences, Taipei 10617, Taiwan}

\author{Mingpu Qin}
\affiliation{Key Laboratory of Artificial Structures and Quantum Control, School of Physics and Astronomy, 
Shanghai Jiao Tong University, Shanghai 200240, China}

\author{ Ulrich Schollw\"{o}ck}
\affiliation{Arnold Sommerfeld Center for Theoretical Physics, Ludwig-Maximilians-Universit\"{a}t M\"{u}nchen, 80333 Munich, Germany}
\affiliation{Munich Center for Quantum Science and Technology (MCQST), 80799 Munich, Germany}

\author{Steven R. White}
\affiliation{Department of Physics and Astronomy, University of California, Irvine, California 92697, USA}

\author{Shiwei Zhang}
\affiliation{Center for Computational Quantum Physics, Flatiron Institute, New York, NY 10010, USA}

\maketitle

\section{Partial particle-hole transformation of the Hubbard model}
When pairing fields are applied in the Hubbard model, the total particle number is not conserved. 
The usual ground-state AFQMC is formulated in the space of Slater determinant with a fixed electron number. While a more general solution is to reformulate AFQMC in 
Hartree-Fock-Bogoliubov (HFB) space \cite{HFB-AFQMC-Shi-2017},
the problem here can be solved without modifying the AFQMC codes, by applying a partial particle-hole transformation \cite{PhysRevX.10.031016} 
\begin{eqnarray}
\hat{c}_{i\uparrow} & \rightarrow & \hat{d}_{i\uparrow}, \quad \hat{c}_{i\uparrow}^{\dagger}  \rightarrow  \hat{d}_{i\uparrow}^{\dagger} \nonumber \\ 
\hat{c}_{i\downarrow} & \rightarrow & \hat{d}_{i\downarrow}^{\dagger}(-1)^{i}, \quad \hat{c}_{i\downarrow}^{\dagger} 
\rightarrow  \hat{d}_{i\downarrow}(-1)^{i}\,, 
\label{eq:partial_ph_trans}
\end{eqnarray}
where $i$ labels the lattice sites in the bipartite lattice.
With this transformation, the $t'$ Hubbard Hamiltonian in Eq.~(1) in the main text turns into
\begin{equation} \label{hubbard_h}
\begin{split}
\hat{H} = -t\sum_{\langle i,j\rangle\sigma}\hat{d}_{i\sigma}^{\dagger}\hat{d}_{j\sigma} - t'\sum_{\langle\langle i,j\rangle\rangle\sigma}s(\sigma)\hat{d}_{i\sigma}^{\dagger}\hat{d}_{j\sigma}  \\
+U\sum_{i}(\hat{m}_{i\uparrow}-\hat{m}_{i\downarrow}\hat{m}_{i\uparrow})
-\mu\sum_{i}(\hat{m}_{i\uparrow} + 1 - \hat{m}_{i\downarrow})\,, 
\end{split}
\end{equation}
where
$s(\uparrow)=+1$ and $s(\downarrow)=-1$,
and $\hat{m}_{i,\sigma}=\hat{d}_{i,\sigma}^{\dagger}\hat{d}_{i,\sigma}$. 
Note that the next near-neighbor
hopping changes sign for down spins after the transformation. The pairing operator $\hat{\Delta}_{ij}\ = (\hat{c}_{i\uparrow}\hat{c}_{j\downarrow}-\hat{c}_{i\downarrow}\hat{c}_{j\uparrow})/\sqrt{2}$ is transformed to: 
\begin{equation}
\hat{\Delta}_{ij}=((-1)^{j+1}\hat{d}_{j\downarrow}^{\dagger}\hat{d}_{i\uparrow}-(-1)^{i}\hat{d}_{i\downarrow}^{\dagger}\hat{d}_{j\uparrow}))/\sqrt{2}
\label{pairing_operator}
\end{equation}
which is now a spin-flip hopping term. The sign of $U$ is
reversed, meaning the interaction turns to attractive. 
Up and down electron now acquire effective chemical potentials $\mu - U$ and $-\mu$, respectively, 
which means $\langle \sum_{i}(\hat{m}_{i\uparrow}\rangle \ne \langle \sum_{i}(\hat{m}_{i\downarrow}\rangle$. 
After the transformation we have
 \begin{equation}
\sum_{i}(\hat{m}_{i\uparrow}+\hat{m}_{i\downarrow})=\sum_{i}(\hat{n}_{i\uparrow}+1-\hat{n}_{i\downarrow})=N_{s},
\end{equation}
 such that
 the total number of electrons equals the number of sites, 
 i.e., the system is at half-filling but
 with spin imbalance.
The random walkers (Slater determinants) are
now represented as $2N \times N_e$ matrix \cite{PhysRevB.94.085103} in the AFQMC calculation, and each orbital in the Slater determinant is now a spin-orbital with a mixture of up and down 
components.

\section{Twist boundary conditions}

Twist boundary conditions (TBC) in $x$-direction means the wave-function satisfies: 
\begin{equation}
\psi ({\mathbf r}_1 + L {\hat e_x}, {\mathbf r}_2, \cdots,  {\mathbf r}_N) =  e^{i\theta_x} \psi({\mathbf r}_1, {\mathbf r}_2, \cdots, {\mathbf r}_N)\,.
\end{equation}
For convenience  we have used $L$ to denote $L_x$, the linear dimension of the periodic cell in 
$x$-direction.
(The definitions of TBC in other directions are similar).
For the two dimensional systems studied in this work, the phases for the two directions are independent of each other, and the phase factors from $x$- and $y$-directions are multiplicative. 
Thus, with no loss of generality, we will  only explicitly write out 
one dimension ($x$) below.
We will assume that the lattice sites are labeled  from $1$ to $L$.

Different gauges can be adopted to realize TBC. 
We discuss two common
choices here. In gauge A, an electron picks up a phase only when it crosses the boundary, while in
gauge B, the phase is split over all bonds evenly. 

\subsection{Gauge A}
In gauge A, if we apply the same twist 
for $\uparrow$ and $\downarrow$-spins in the repulsive Hubbard model, we have
\begin{eqnarray}
c^{\dagger}_{L+1,\sigma} &=& \exp(i\theta) c^{\dagger}_{1,\sigma}  \nonumber \\
c_{L+1,\sigma} &=& \exp(-i\theta) c_{1,\sigma}
\end{eqnarray}
So the hopping between the last and first site is modified as
\begin{equation}
-t\hat{c}_{1\sigma}^{\dagger}\hat{c}_{L\sigma}+h.c \rightarrow 
-t\exp(i\theta)\hat{c}_{1\sigma}^{\dagger}\hat{c}_{L\sigma}+h.c
\label{phase_1}
\end{equation}
while the other hopping terms remains unchanged. After the partial particle-hole transformation in Eq.~(\ref{eq:partial_ph_trans}), the spin up term is unchanged, but the phase for down spin changes to $-\theta$ as
\begin{equation}
-t\exp(-i\theta)\hat{d}_{1\downarrow}^{\dagger}\hat{d}_{L\downarrow}+h.c
\label{phase_2}
\end{equation}
The same is true for the $t'$ term.

With TBC,
we also need to modify the definition of the pairing operator ($\hat{\Delta}_{kj}\ = (\hat{c}_{k\uparrow}\hat{c}_{j\downarrow}-\hat{c}_{k\downarrow}\hat{c}_{j\uparrow})/\sqrt{2}$, for
bonds connecting nearest-neighbor sites, $\langle jk\rangle$) for the bond connecting the first and last site as
\begin{equation}
\hat{\Delta}_{1L}=\exp(-i\theta)(\hat{c}_{1\uparrow}\hat{c}_{L\downarrow}-\hat{c}_{1\downarrow}\hat{c}_{L\uparrow})/\sqrt{2}\,.
\label{phase_pairing}
\end{equation}
When applying the pairing 
field to 
calculate the pairing order, we need to include the phase in Eq.~(\ref{phase_pairing}) when twist boundary conditions are imposed. In the AFQMC calculation, the pairing operator in Eq.~(\ref{phase_pairing}) can be transformed, following Eq.~(\ref{eq:partial_ph_trans}), as
\begin{equation}
\hat{\Delta}_{1L}=\exp(-i\theta)(\hat{d}_{1\downarrow}^\dagger\hat{d}_{L\uparrow}-(-1)^L\hat{d}_{L\downarrow}^\dagger\hat{d}_{1\uparrow})/\sqrt{2}
\label{phase_pairing-U}
\end{equation}
Other pairing terms are transformed to the $-U$ case following Eq.~(\ref{pairing_operator})

\subsection{Gauge B}

We next consider gauge B in a similar setup to Gauge A. 
Now the phase is spread evenly over 
each bond and we have, for the repulsive model
\begin{eqnarray}
c_{j\sigma}^{\dagger}&\rightarrow& c_{j\sigma}^{\dagger}\exp(i(j-1)\frac{\theta}{L}) \nonumber \\ c_{j\sigma}&\rightarrow& c_{j\sigma}\exp(-i(j-1)\frac{\theta}{L})  
\end{eqnarray}
and
\begin{eqnarray}
c^{\dagger}_{L+1,\sigma} &=& \exp(i\theta) c^{\dagger}_{1,\sigma}  \nonumber \\
c_{L+1,\sigma} &=& \exp(-i\theta) c_{1,\sigma}
\end{eqnarray}
The nearest neighbor hopping term is then modified to 
\begin{equation}
-t\sum_{j}\exp(i\theta/L) \hat{c}_{j+1\sigma}^{\dagger}\hat{c}_{j\sigma} + h.c.
\end{equation}
The $t'$ term has similar form. 

The pairing operator $\hat{\Delta}_{kj}\ = (\hat{c}_{k\uparrow}\hat{c}_{j\downarrow}-\hat{c}_{k\downarrow}\hat{c}_{j\uparrow})/\sqrt{2}$ is modified as
\begin{equation}
  \hat{\Delta}_{kj}\ = (\hat{c}_{k\uparrow}\hat{c}_{j\downarrow}-\hat{c}_{k\downarrow}\hat{c}_{j\uparrow})\exp(-i(k+j-2)\theta/L)/\sqrt{2}
   \label{pair-2}
\end{equation}
For the bond connecting the first and last site, we have
\begin{equation}
  \hat{\Delta}_{L,1}\ = (\hat{c}_{L\uparrow}\hat{c}_{1\downarrow}-\hat{c}_{1\downarrow}\hat{c}_{L\uparrow})\exp(-i(2L-1)\theta/L)/\sqrt{2}
   \label{pair-2-b}
\end{equation}
We can then follow the particle-hole transformation in Eq.~(\ref{eq:partial_ph_trans}) 
to transform the definition of pairing order to the negative $U$ model, which is used in the AFQMC calculation.

\subsection{The equivalence of the two gauges}
The two gauges discussed above are equivalent and physical quantities should have the same values under them, 
which we have explicitly verified.
Since the interaction term is independent of the twist angle, it is convenient
to test the TBC implementation in 
non-interacting systems.
For example, 
in a $20\times4$ lattice with $t'=-0.2t$, $\mu=0.8$, and twist angle $\theta_x = 1.2994\pi, \theta_y = 0.6026\pi $, 
it is easily checked in all our codes  
that physical quantities, such as the energy per site ($-1.15861112$), average pairing order per bond ($0.01108939$), and the electron density ($0.82422077$), are all exactly the same under the two gauges. 

\section{Self-consistent constraint in
AFQMC} 
We apply magnetic and pairing pinning fields to probe the corresponding response in the studied systems. A self-consistent 
procedure in AFQMC allows us to apply a constraint to remove the sign problem. 
We describe the pinning field calculations and the self-consistency procedure below.

The magnetic pinning fields are typically 
applied in cylindrical cells, to one or both ends of the cylinder, not 
in the rest of the cell. 
We try different configurations of the magnetic pinning fields to probe the possible magnetic order or correlation.
The strength of the fields is fixed at $h_m = 0.25$, and limited to only the 
edge(s) of the cylinder.
For most of the systems, we applied anti-ferromagnetic pinning fields at the open edges of the studied cylinders ($(-1)^{(i_x + i_y)}h_m$ for $i_x = 1$ and $L_x$). In some cases, we also 
test a pinning field configuration with 
a $\pi$ phase to the pinning magnetic fields on the right edge as ($(-1)^{(i_x + i_y)}h_m$ for $i_x = 1$ and $(-1)^{(i_x + i_y + 1)}h_m$ for $i_x = L_x$), and compare the energies to determine the true ground state, the one with the lower energy. Note that it is important in this scheme to examine progressively larger (longer) systems, in order to remove the 
effect of the local pinning field.
Ref.~\cite{PhysRevResearch.4.013239} includes further details of our analysis 
method and how we extract information in the TDL.

\begin{figure}[t]
\centering{
	\includegraphics[width=0.98\linewidth] {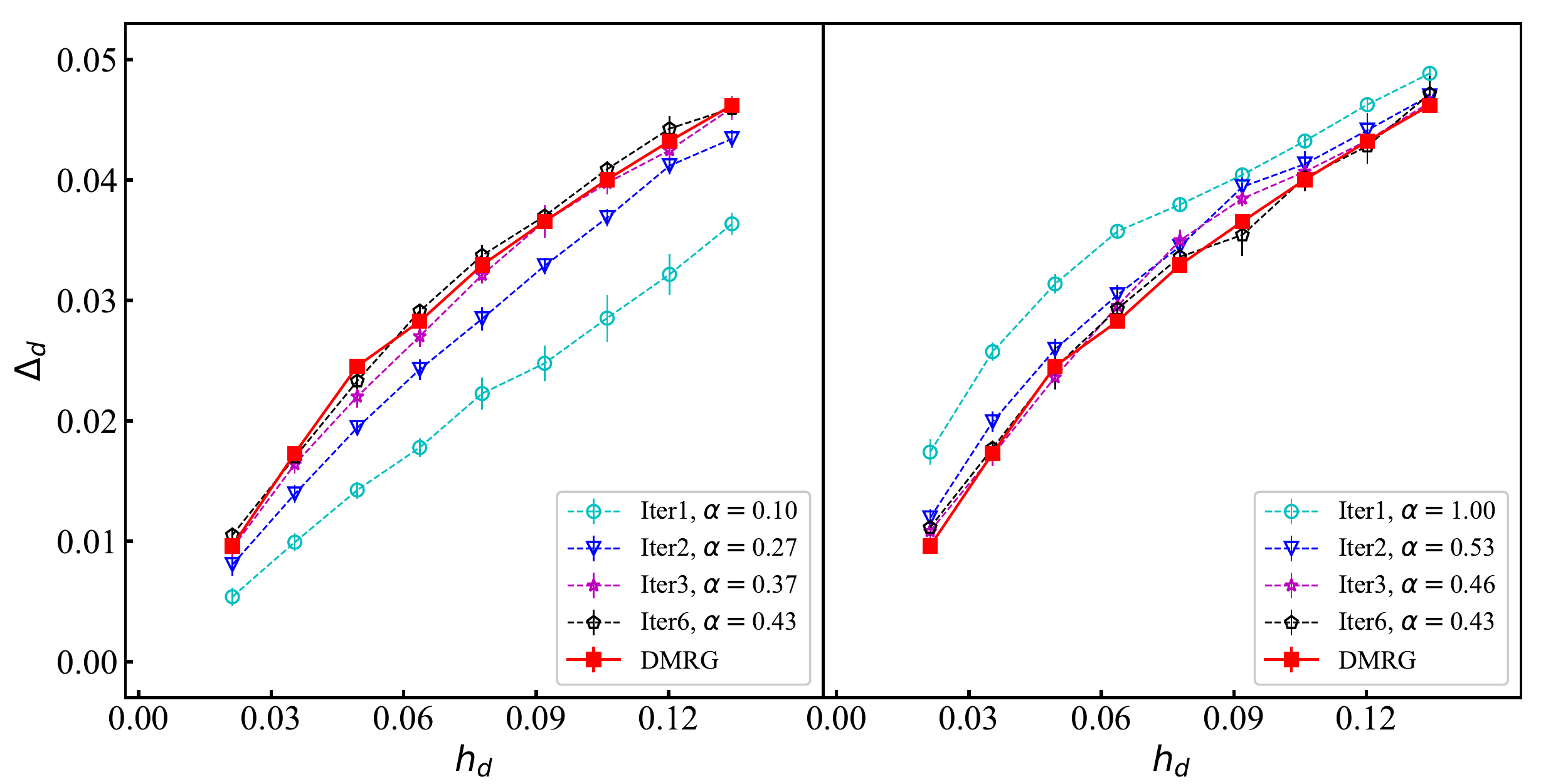}
 }
	\caption{ Robustness of the self-consistent constraint in AFQMC. 
     An example is shown for
     the pairing order parameter 
     in the $1/4$ hole-doped Hubbard model on a $16 \times 4$ cylinder. 
     On the left panel, the self-consistent calculation starts with an initial value of 
     $\alpha = 0.1$. The calculation converges with a handful or iterations 
     to the exact result (DMRG, red) in this system, with a converged value
     $\alpha = 0.43$.
     On the right panel, the calculation is initialized with $\alpha=1.0$, and converges from the opposite direction to the same result. 
     }
	\label{iteration_example}
\end{figure}

To compute the pairing order parameter, 
we apply global pairing fields 
across the entire simulation cell,
similar to Ref.~\cite{PhysRevX.10.031016}.
To probe the $d$-wave pairing response
the applied pair-inducing fields 
on vertical and horizontal bonds have the same strength $h_d$ but opposite signs.
The Hamiltonian with pairing fields is
\begin{equation}
\hat{H}'(h_d) = \hat{H} +  \hat{H_d}(h_d)
\end{equation}
where 
\begin{equation}
\hat{H_d}(h_d) = -h_d \sum_{\langle i,j\rangle} b_{ij} \frac{\hat{\Delta}_{ij} + \hat{\Delta}^{\dagger}_{ij}}{2}\,, 
\end{equation}
where $b_{ij}=+1$ for 
a bond connecting two nearest-neighbors $i$ and $j$ in the $x$-direction and  $b_{ij}=-1$ if $\langle ij\rangle$ is in the $y$-direction.

The pairing order is calculated from the derivative of the ground state energy $E'(h_d)$ with respect to $h_d$, following the Hellmann-Feynman theorem.
Recall these calculations are performed 
in the particle-hole-transformed 
attractive Hubbard Hamiltonian in 
Eq.~(\ref{hubbard_h}).
We take the ground state of the non-interacting Hamiltonian ($U=0$ in Eq.~(\ref{hubbard_h})) with the pairing field $\alpha h_d$ as trial wave function,
where $\alpha$ is a parameter to be determined in the self-consistent iterations.  In the first step, we choose an arbitrary $\alpha$ and calculate the average value of pairing order with AFQMC
using the corresponding trial wave function as a constraint. We then tune the value of $\alpha$ 
by minimizing the difference between the 
pairing orders given by the
non-interacting wave-function
 and the previous iteration of AFQMC.
We then carry out the next iteration AFQMC calculation with the new trial wave-function. This process is repeated until $\alpha$ 
is converged. 
In each mean-field solution we tune the value of chemical potential to target the desired spin imbalance 
(i.e., the electron density in the 
repulsive model). 
In performing TABC, we determine the final 
value of $\alpha$ via averaging over different twist angles. The value is 
found to converge quickly, so a small 
set of pilot calculations can be performed 
first to obtain a good estimate. More computations can be added if further precision is needed.

In Fig.~\ref{iteration_example}, we show the computed pairing order in the self-consistent process for the $1/4$ hole doped $t'$ Hubbard model on a $16 \times 4$ cylinder. DMRG results are also shown, because for this narrow system, 
it provides a reference result which is essentially exact. 
In Fig.~\ref{iteration_example}, we start the self-consistent calculation with an initial value $\alpha = 0.1$ (the left panel). After 6 iterations,
the pairing order converges to the DMRG results. The converged value of $\alpha$ is $\alpha = 0.43$.
We also obtain the same converged pairing order and $\alpha$ value by starting the self-consistent process with
$\alpha = 1.0$ (the right panel), indicating the self-consistent calculation is independent of the initial value of $\alpha$.

\section{Sensitivity of order to system sizes and boundary conditions}
\begin{figure}[t]
	\includegraphics[width=0.98\linewidth]{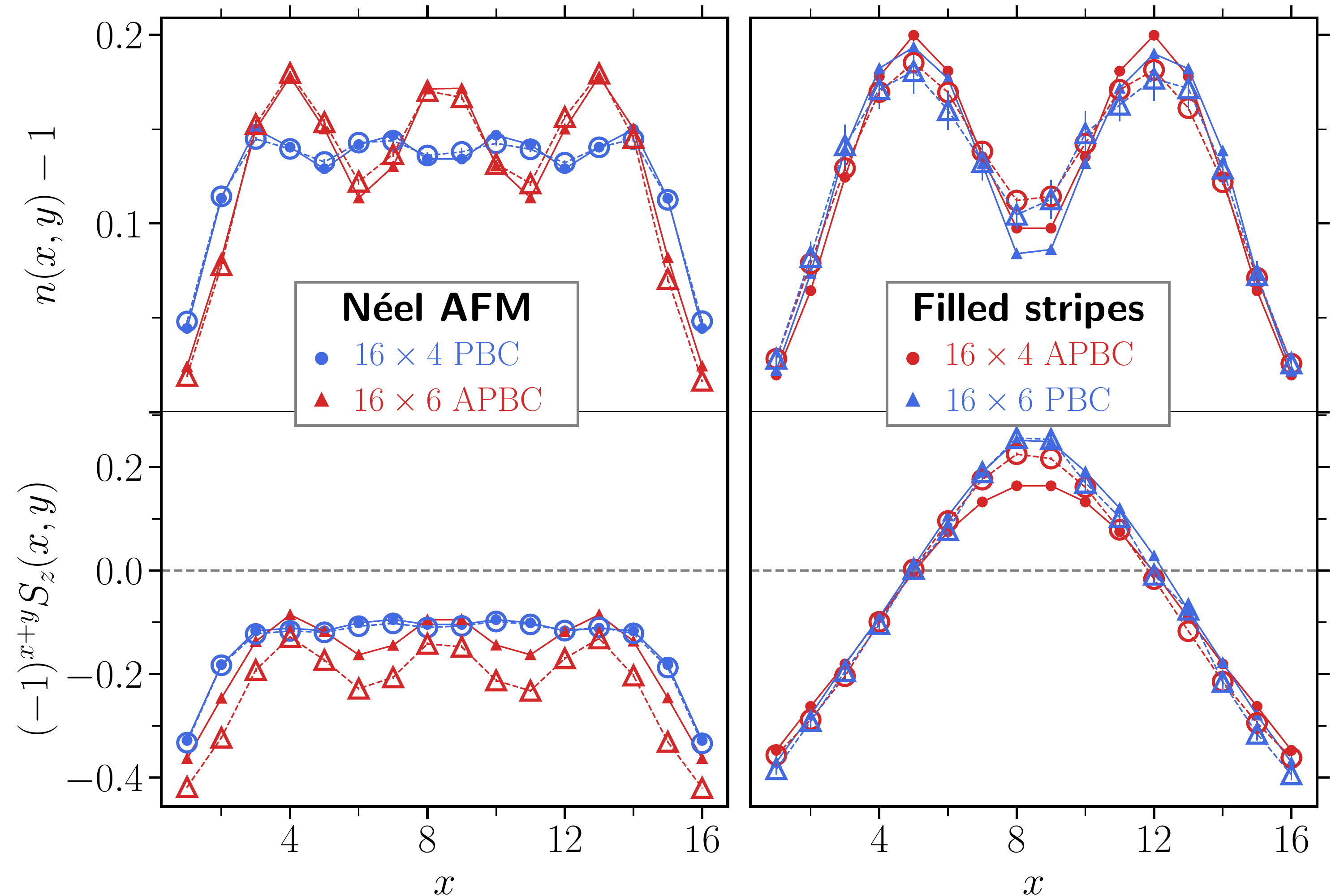}        
	\caption{ Strong sensitivity of the spin and charge orders to system sizes and boundary conditions. Here we show results for four different systems, all with the same bulk parameters ($\delta=1/8$, electron doped), but different in size and boundary conditions. Line cuts of the doped electron density (top panels) and staggered spin density (bottom) are shown, with combinations of two system sizes, $16\times4$ and $16\times6$, and periodic (PBC) and antiperiodic (APBC) boundary conditions. Antiferromagnetic pinning fields have been applied at the left and right edges of the open cylinders. 
 Two distinct types of states appear, N\'eel AFM with fairly uniform electron densities (left panels), and filled stripe states (right panels). 
 Good agreement is found between DMRG (filled symbols) and AFQMC (empty symbols); the differences are tied to the sizes and boundary conditions. 
   } 
	\label{size-effect}
\end{figure} 
In Fig.~\ref{size-effect} we show an example of the strong sensitivity of the ground states to system sizes and boundary conditions (BCs).  Two entirely different ground states are obtained for the same physical parameters from four different combinations of size/BCs, 
with an alternation between the effects of size versus BC. 
Consistent results are seen from both methods. 
This sort of sensitivity is also observed on the hole-doped side (see Fig.~\ref{stripe-hole}). 
To determine the order in the thermodynamic limit in these systems thus requires computations in significantly larger sizes than 
has been previously reached.
Below, we also show the presence of numerous low-lying states whose ordering in energy can be affected by size and BCs. In many cases these low-lying states can be tied to different stripe configurations.  In the case of Fig.~\ref{size-effect}, the system may be close to a phase boundary between the two types of states\cite{doi:10.1073/pnas.2109978118}. 
The twist-averaging procedure adopted here tends to average over the various states, which allows better extrapolation to the TDL 
compared to earlier approaches\cite{2019Sci...365.1424J}.

\section{Level crossing with applied pairing fields}

\begin{figure}[t]
\centering{
\includegraphics[width=0.49\linewidth]{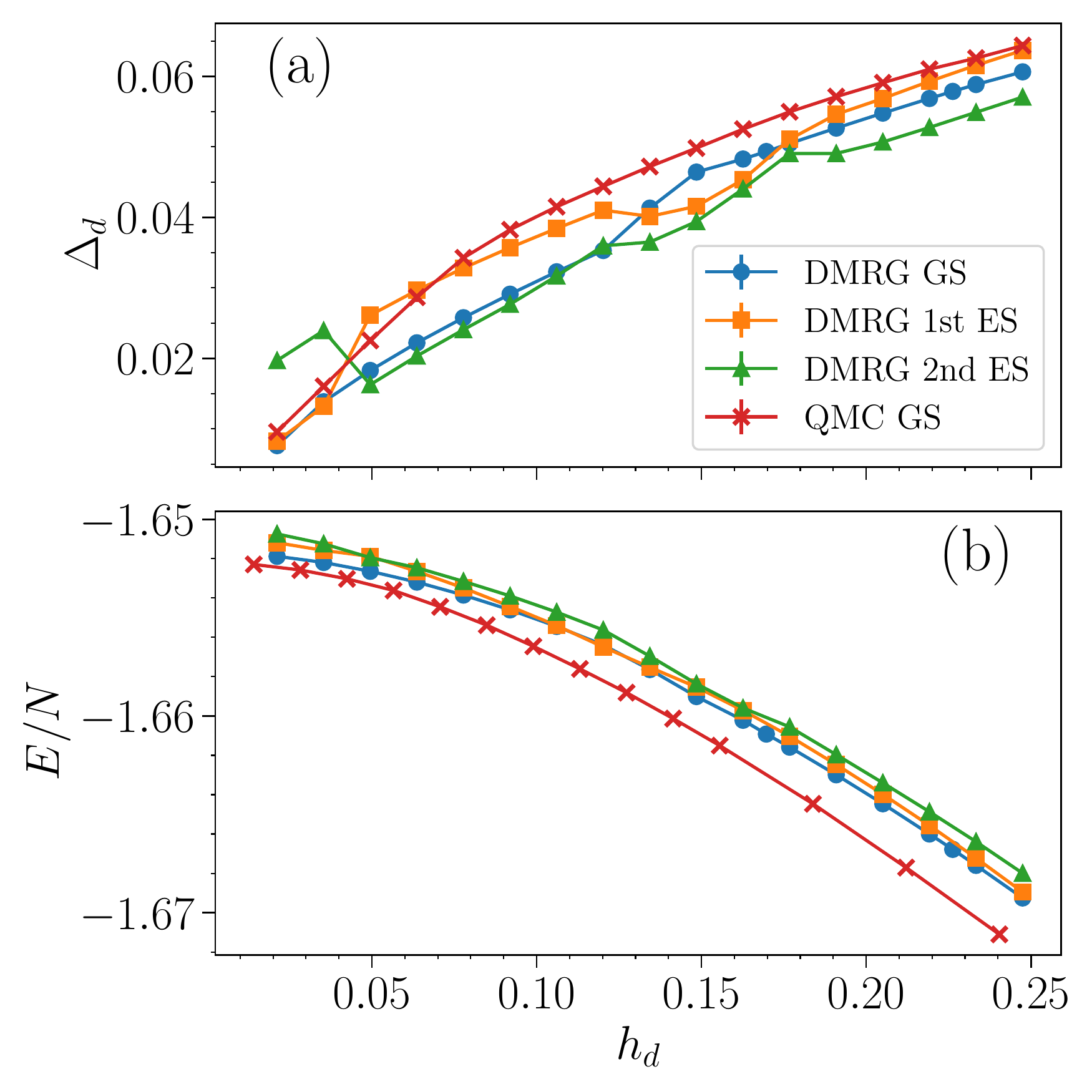}
\includegraphics[width=0.49\linewidth]{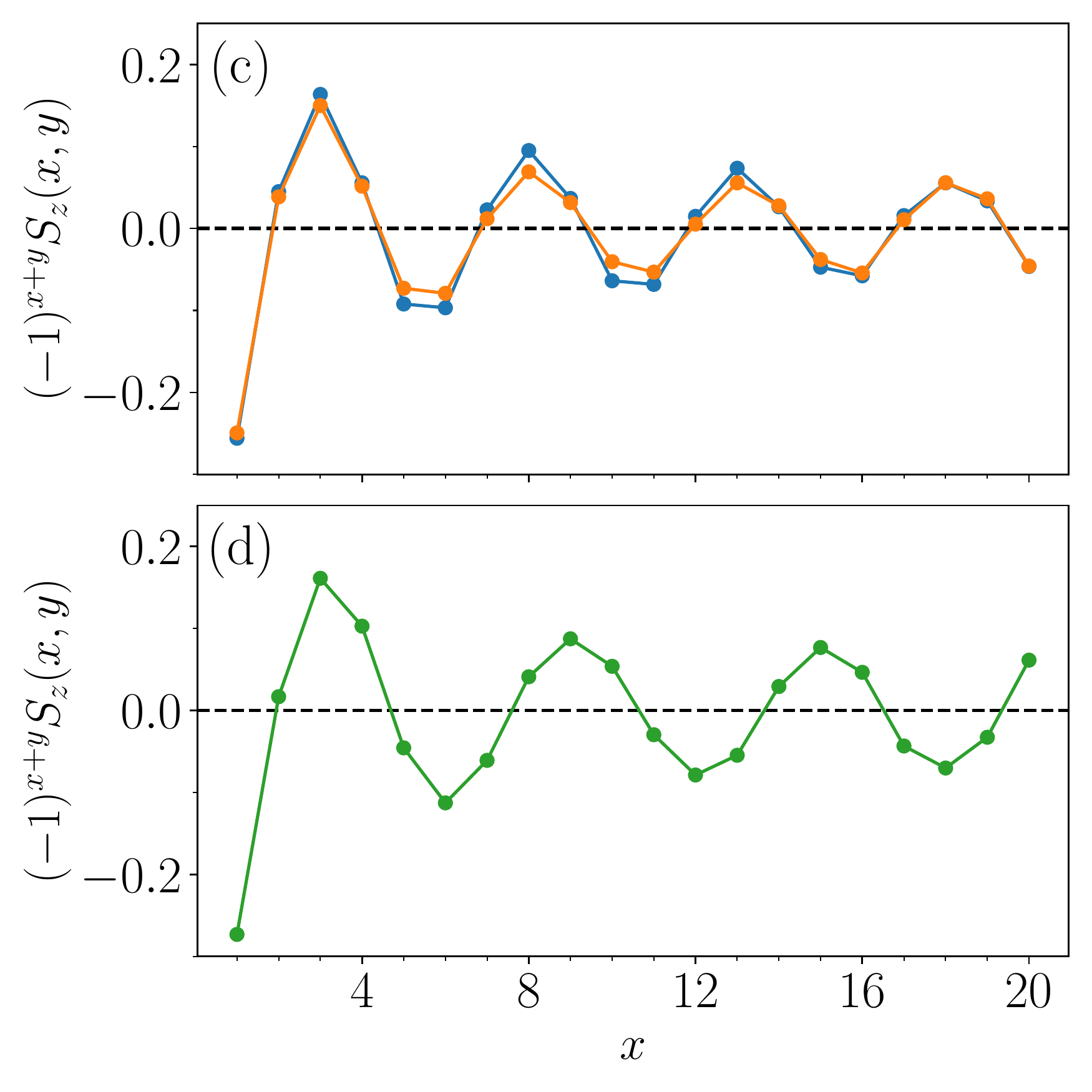}
}
\caption{
Illustration of low-lying states and level-crossing.
The pairing order $\Delta_d$ (a) and energy per site (b) are shown as functions of the global pairing field strength $h_d$ for a $1/5$ hole doped system in a $20\times 4$ cylinder. The error bars are smaller than the symbol sizes.
In (c) and (d) the staggered spin densities are shown along the $x$ direction for the three low-lying states from DMRG, for $h_d\approx0.0075$, the smallest $h_d$ we consider. The AF magnetic pinning fields are applied at the left open edge.
} 
\label{fig:low_lying}
\end{figure} 

 In addition to the enhanced sensitivity of the ground state to the boundary conditions and system sizes in the $t'$ Hubbard model,
 the evolution of the ground state with the strength of the applied pairing field 
 is subtle, and
 creates another computational challenge. 
In Fig.~\ref{fig:low_lying} (a) and (b), as an example, we show the evolution of a few low-lying states as a function of 
$h_d$,
in a $20 \times 4$ system at $1/5$ hole doping.
As $h_d$ decreases, the low-lying states,
which are separated by tiny energy differences (note the small energy scale in b),
exhibit crossovers between several branches.
The different branches are characterized by different numbers of stripes in the states, as shown in Fig.~\ref{fig:low_lying} (c) and (d).
That $d$-wave pairing field induces strong level crossings 
is another indication of the intimate connection between the fluctuation of the stripe state and superconductivity in the system.
As mentioned, such level-crossings 
make the comparisons between DMRG and AFQMC more challenging,
as each calculation is often sensitive to 
even small variations in the calculational parameters. 
However this effect is reduced by employing twist averaging, in which all the low-lying states are sampled. As can be seen 
in Fig.~5 
in the main text, 
TABC effectively treats the crossovers 
as a function of $h_d$, which results
in smooth curves, and  
DMRG and AFQMC agree very well. 

\section{Supplemental data}

\subsection{Hole density for $1/8$ hole doped systems}
\begin{figure}[t]
        \centering{
	\includegraphics[width=0.98\linewidth]{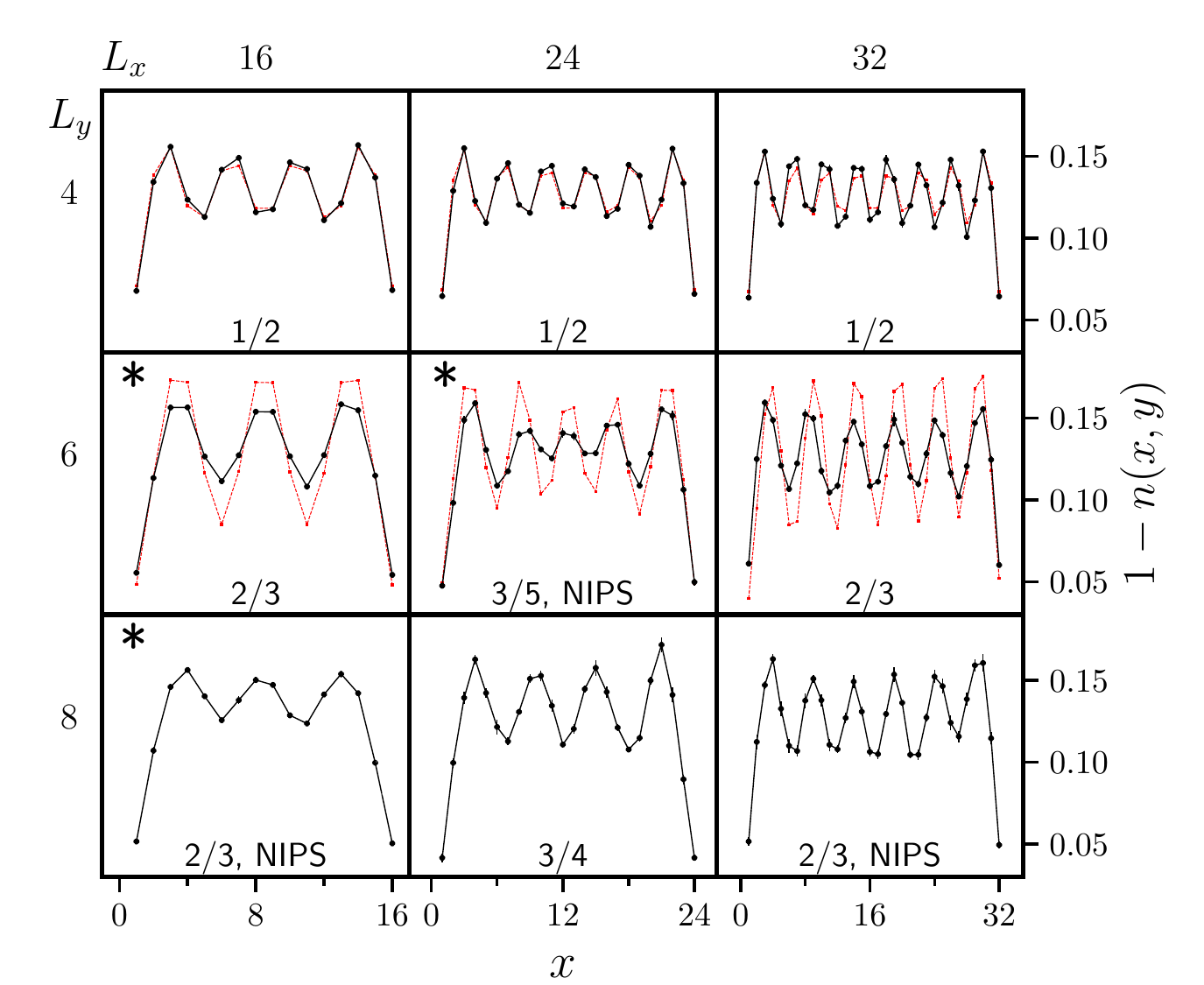}
        }
	\caption{
 Hole density for the systems in Fig.~2 in the main text for $1/8$ hole doping. 
 The evolution of the stripe patterns is shown versus system size.
 The hole densities are shown as linecuts 
 along the length of the cylinders. 
 The length of the cylinder ($L_x$) is varied across 
 the three columns 
and the width ($L_y$) across rows. 
AFM pinning fields are applied at the two edges of the cylinder ($x=1$ and $x=L_x$), either in phase or with a $\pi$-phase shift
(marked by an asterisk);
the one with lower energy is shown.
The filling fraction $f$ of each stripe pattern is indicated, with NIPS denoting non integer-pair stripes. 
 DMRG results (red) are shown for width-4 and 6 systems, and AFQMC (black) are in good agreement with them.
} 
\label{stripe-hole}
\end{figure} 
In Fig.~\ref{stripe-hole}, we show the hole density for the systems in Fig.~2 in the main text. For systems with width 4 and 6, for which DMRG is available, 
we find good agreement between AFQMC amd DMRG results.
In width-6 systems, the discrepancies are somewhat larger in the density here compared to the spin 
in Fig.~2 in the main text. 
This is likely because of a combination of two factors. First AFQMC has shown in 
$t'=0$ Hubbard model a slight tendency to under-estimate the amplitude of the density fluctuations in 
stripes \citep{PhysRevResearch.4.013239, PhysRevB.94.235119}. Second, in some cases we have seen indications that the DMRG may not have reached full convergence in width-6 systems, even with the very large bond dimensions we were able to do.

\subsection{Additional data on pairing order parameter}
In this subsection, we include  
the finite size data for the pairing order, as well as the extrapolation process to obtain the spontaneous pairing order in the thermodynamic limit
which is plotted in Fig.~1 in the main text.

\subsubsection{Electron doped region}

\begin{figure*}[t]
        \centering{
	\includegraphics[width=0.32\linewidth]{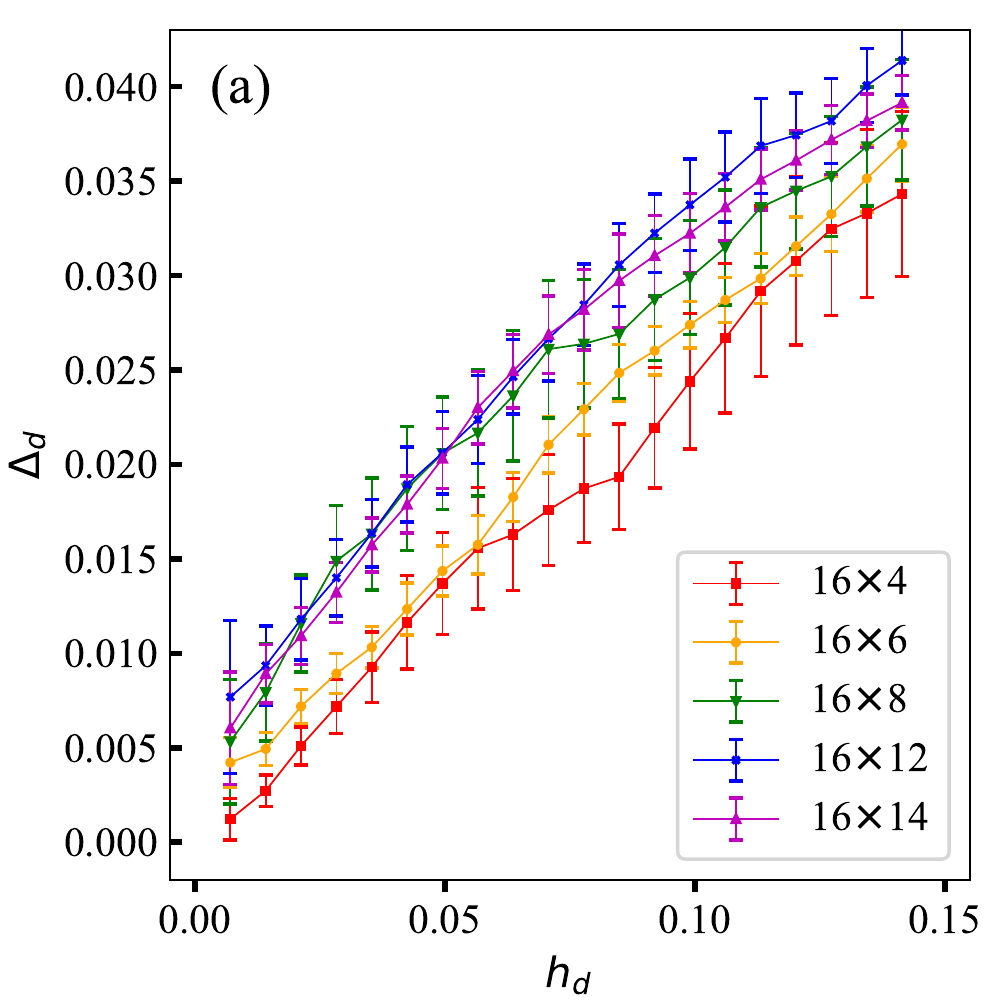}
        \includegraphics[width=0.32\linewidth]{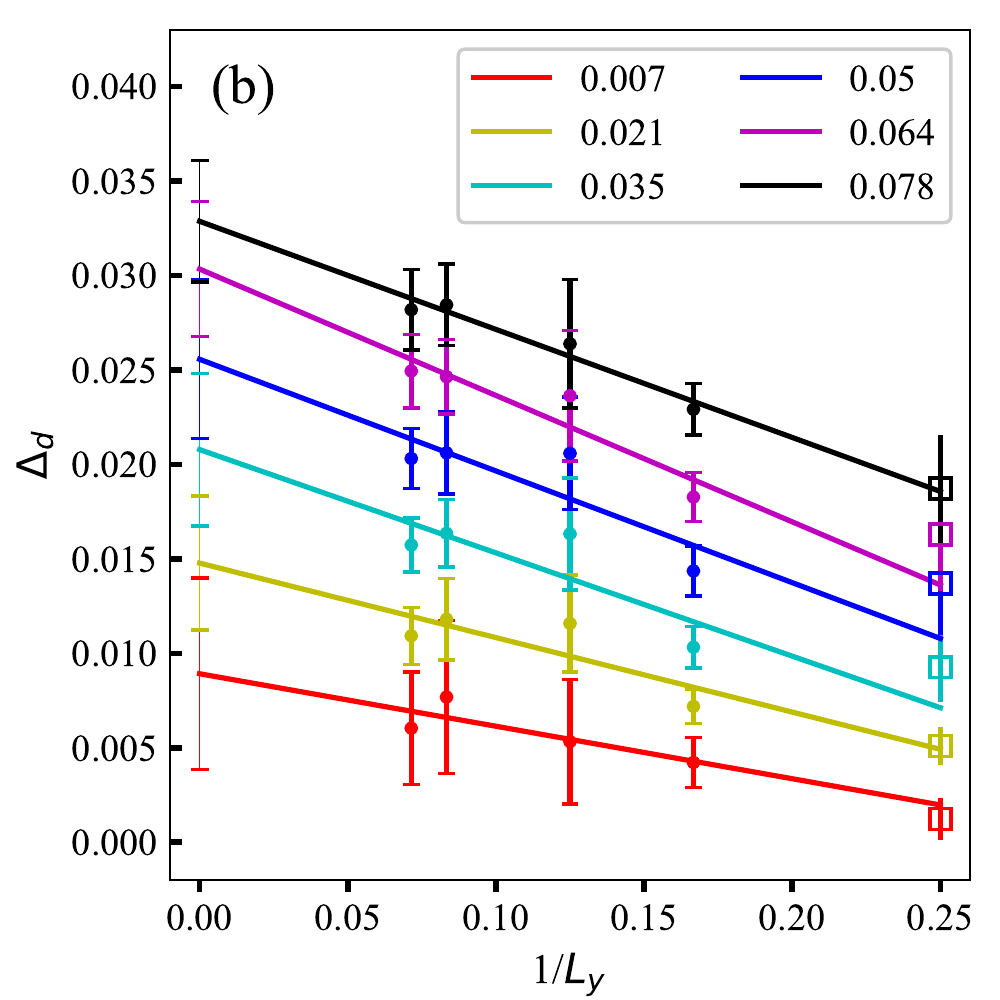}
        \includegraphics[width=0.32\linewidth]{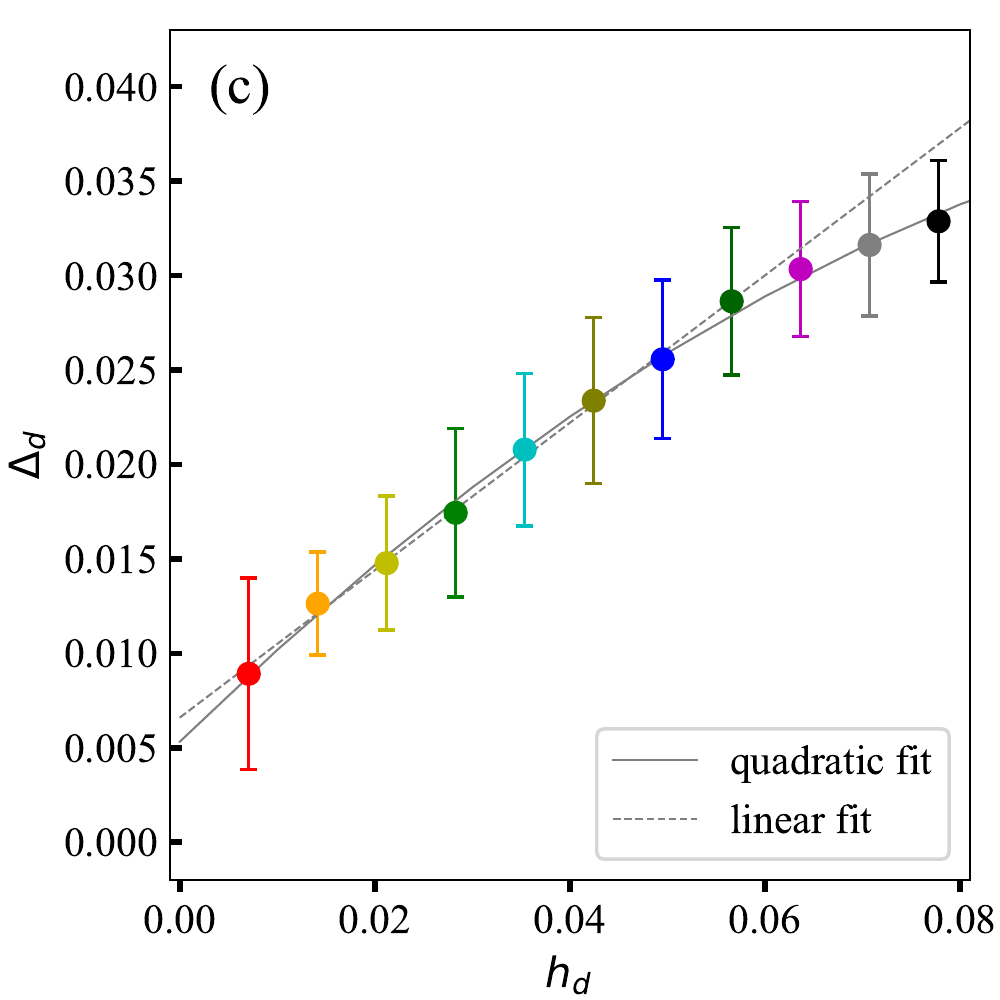}
	}
        \caption{The extrapolation of pairing order for 1/8 electron doping. 
        (a) shows the pairing order
parameter versus pairing field for different system sizes. 
The data for each torus of $L_x\times L_y$ is obtained from TABC with quasi-random 
twists $(k_x,k_y)$. The number of twists
is a few dozens for smaller systems and
about a dozen 
for the larger systems. 
(b) shows extrapolation of the pairing order with the
width of the system for each fixed pairing field. In (c) the extrapolated value in (b) is plotted against
pairing field and then an extrapolation of the pairing field to $h_d\rightarrow 0$ gives the spontaneous pairing order at
thermodynamic limit. Both linear and quadratic fits give consistent results.
} 
	\label{pair-extra-1-8-ele}
\end{figure*}

\begin{figure*}[t]
        \centering{
	\includegraphics[width=0.32\linewidth]{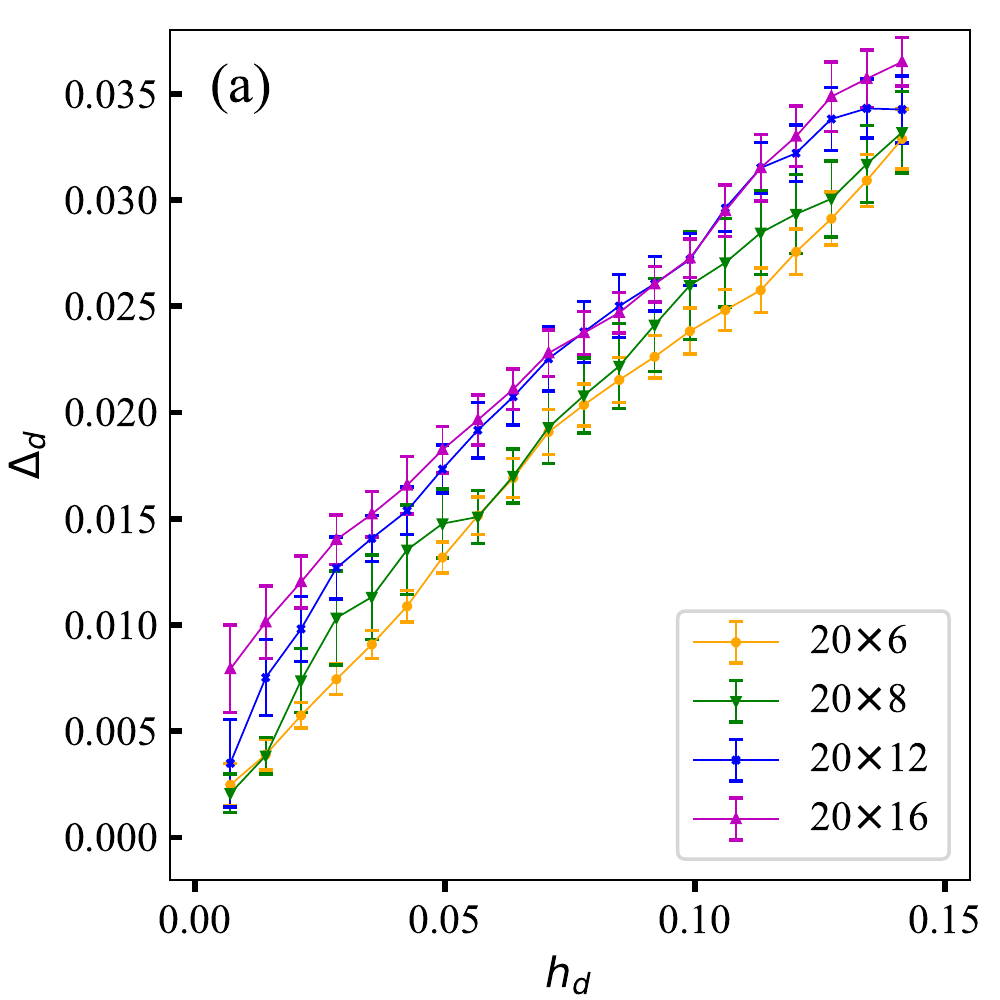}
        \includegraphics[width=0.32\linewidth]{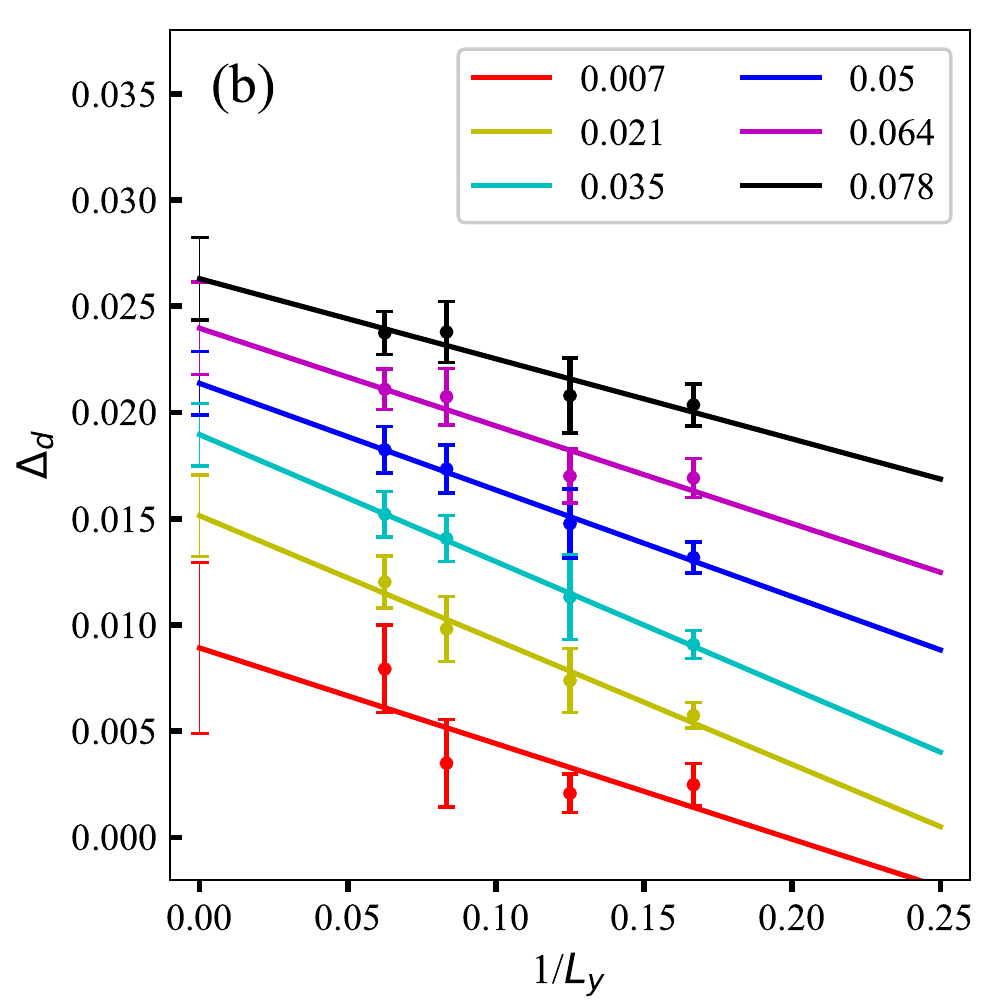}
        \includegraphics[width=0.32\linewidth]{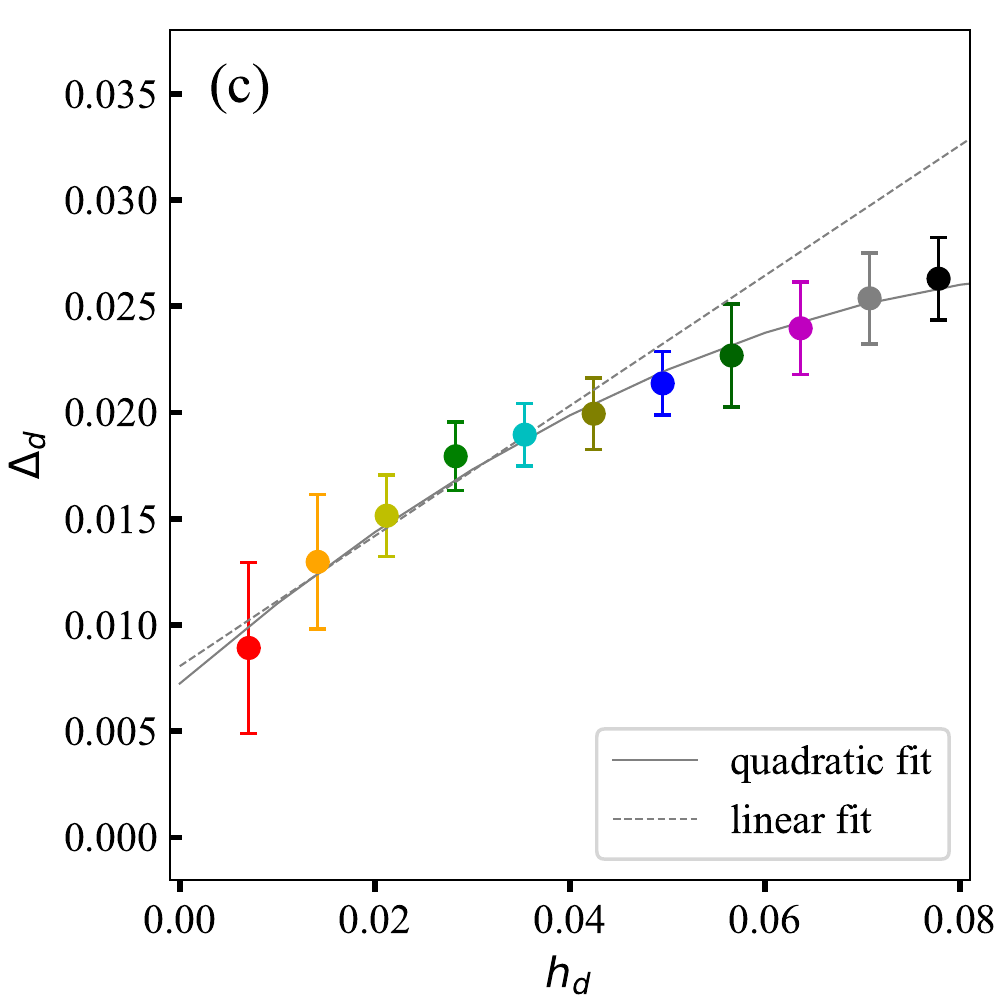}
	}
        \caption{Similar as Fig.~\ref{pair-extra-1-8-ele} but for $1/5$ electron doping.} 
	\label{pair-extra-1-5-ele}
\end{figure*}

\begin{figure*}[t]
        \centering{
	\includegraphics[width=0.32\linewidth]{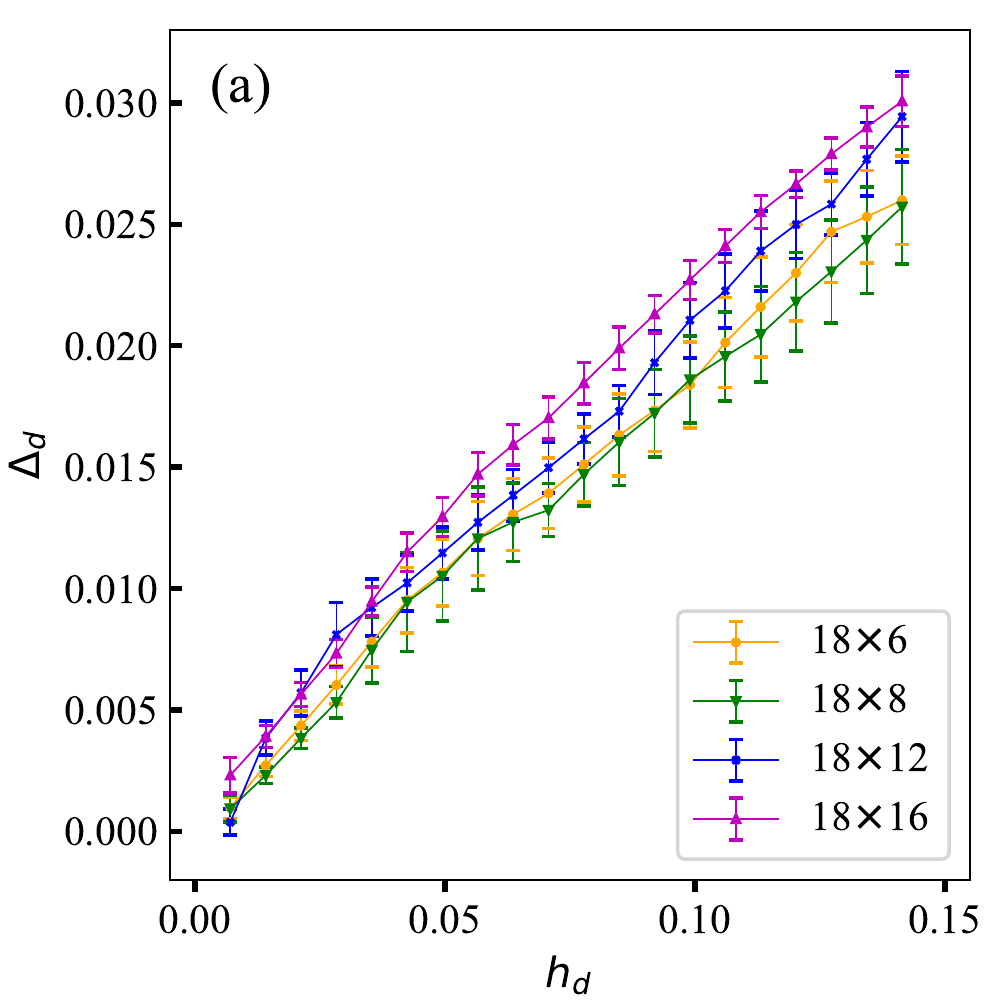}
        \includegraphics[width=0.32\linewidth]{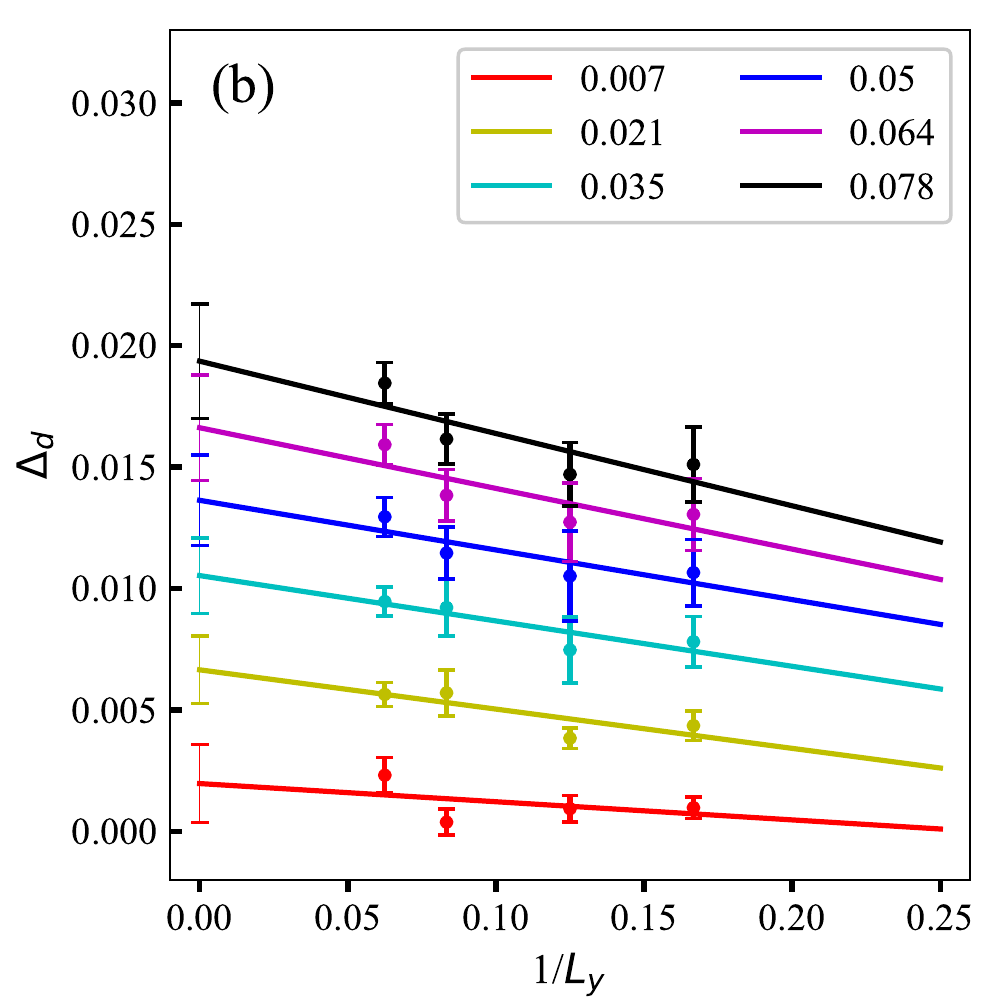}
        \includegraphics[width=0.32\linewidth]{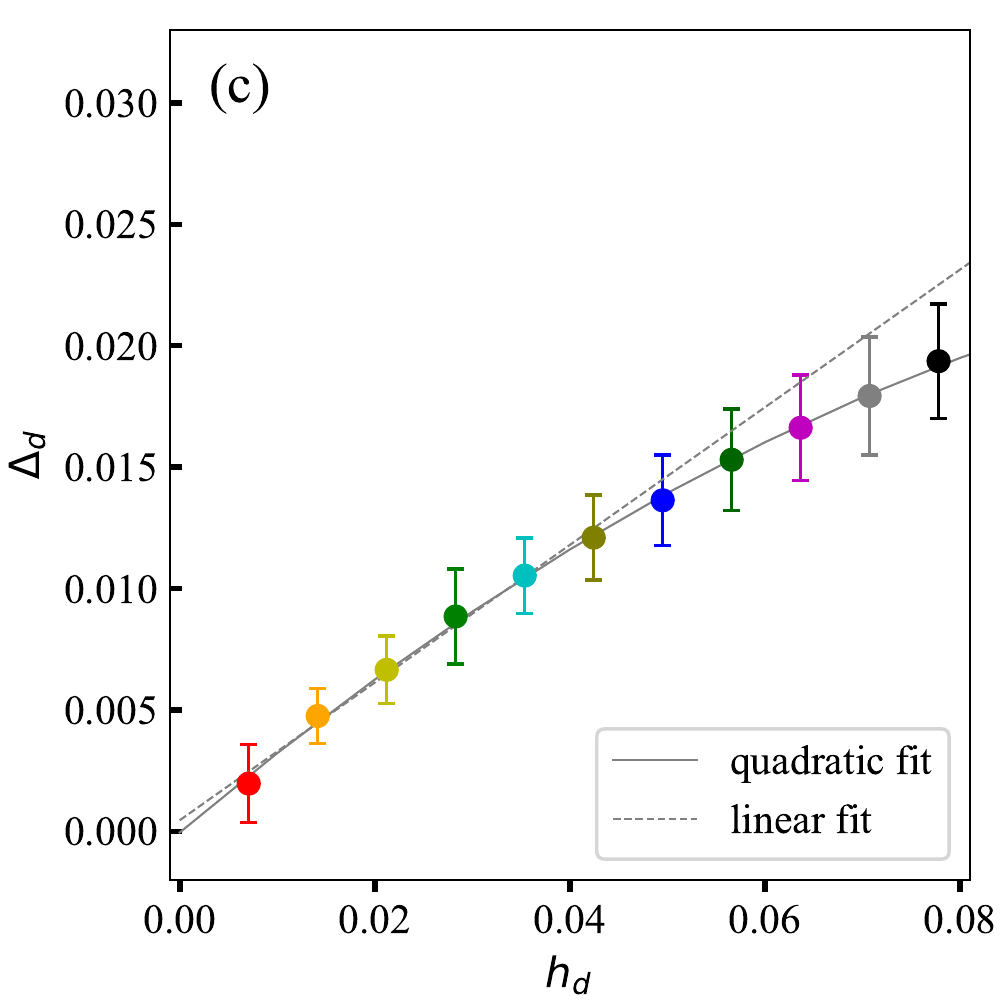}
	}
        \caption{Similar as Fig.~\ref{pair-extra-1-8-ele} but for $1/3$ electron doping.} 
	\label{pair-extra-1-3-ele}
\end{figure*}

In 
Figs.~\ref{pair-extra-1-8-ele}, \ref{pair-extra-1-5-ele}, and \ref{pair-extra-1-3-ele}
we present the data 
for pairing order in the electron-doped region, with $\delta = 1/8, 1/5$, and $1/3$ respectively. 
As discussed in the main text, 
we use fully periodic systems (i.e., 
torus simulations cells of $L_x\times L_y$) and perform TABC with quasi-random 
twists. We have verified that the 
results have converged with respect to 
$L_x$ to within our statistical error. 
When extrapolating the pairing orders with width of the system, we omit width-4 systems.
When extrapolating the TDL values versus $h_d$, we perform both linear and quadratic fits. 
In the linear fits, we use $h_d<0.05$,
when the data is clearly in the linear response regime. In the quadratic fits,
we include more data points,  
10-12 values of $h_d$. 
We find that the
resulting extrapolated $\Delta_d(h_d\rightarrow 0)$ values are often indistinguishable.
When there is a difference, the result from the better fit (smaller $\chi^2$) is used.
The final pairing order at TDL is $0.007(3),0.007(4)$, and $0.000(2)$ for $\delta = 1/8, 1/5$, and $1/3$, respectively,
as reported in Fig.~1 in the main text.

\subsubsection{Hole doped region}

\begin{figure*}[t]
        \centering{
	\includegraphics[width=0.32\linewidth]{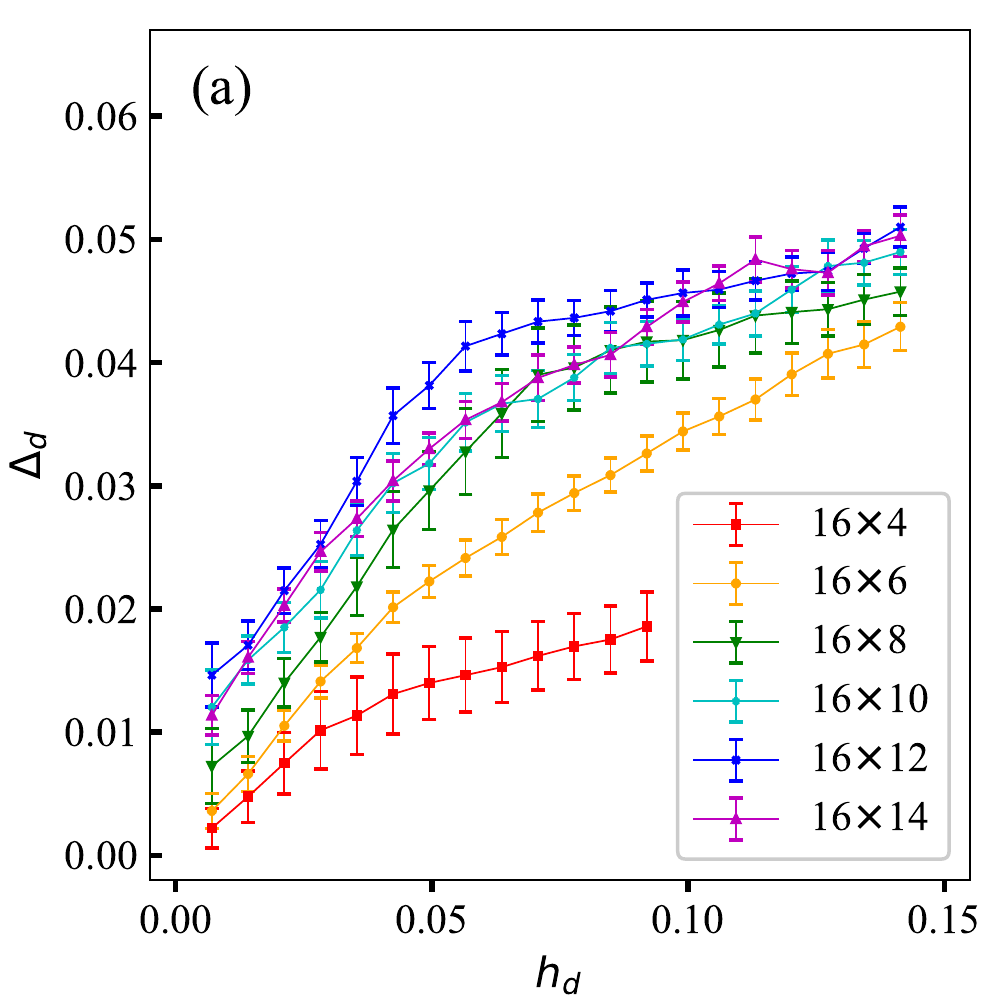}
        \includegraphics[width=0.32\linewidth]{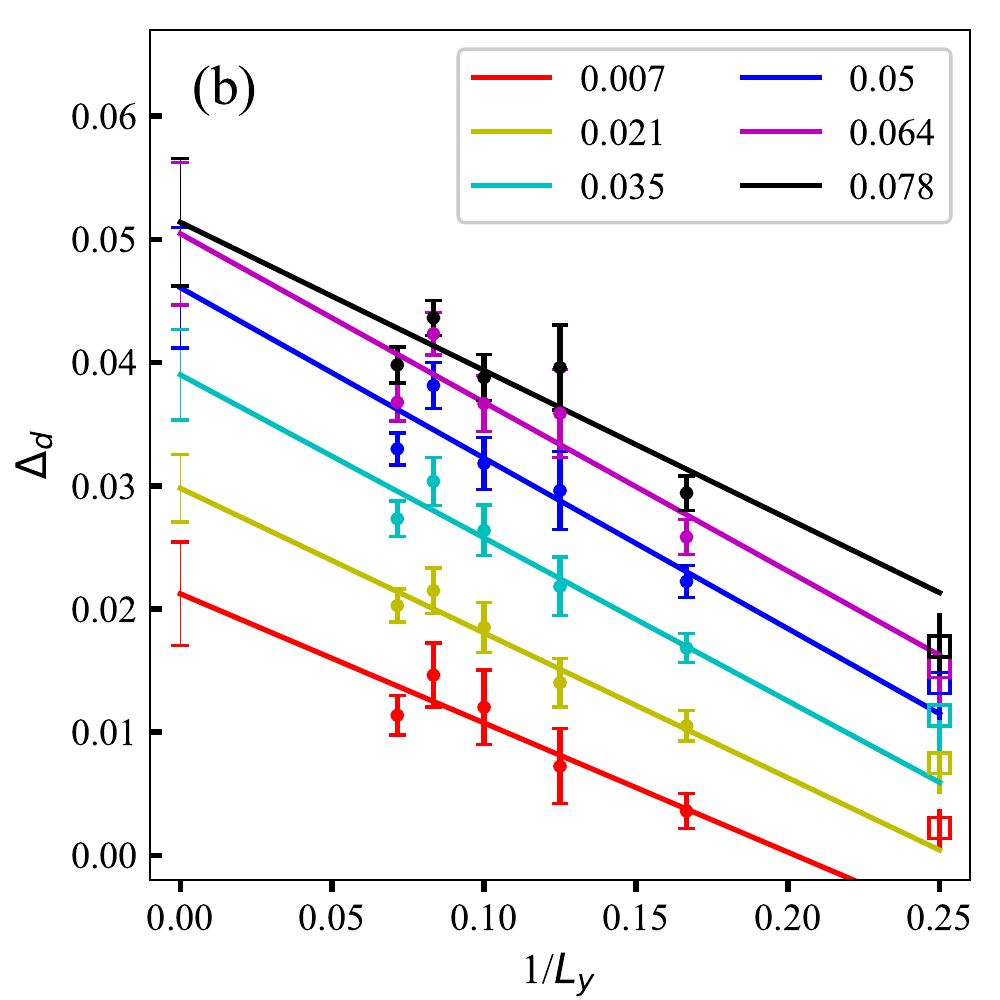}
        \includegraphics[width=0.32\linewidth]{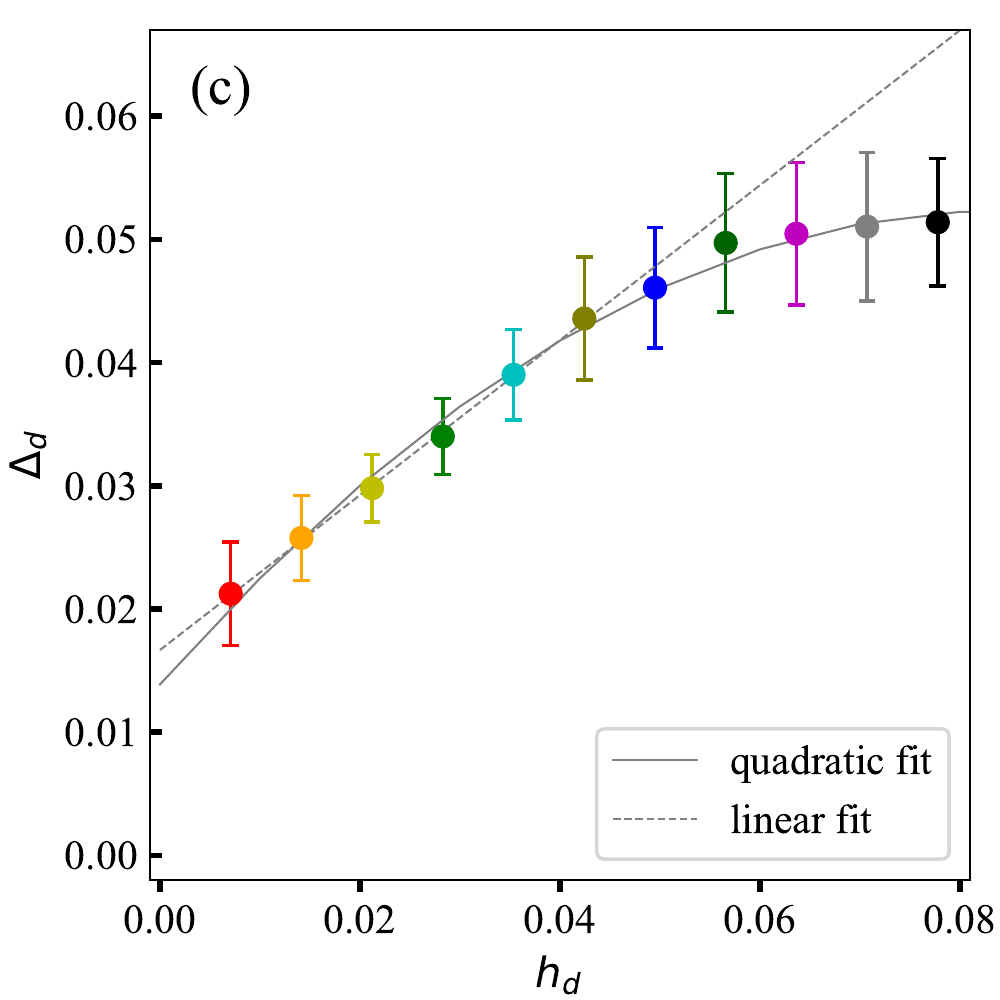}
	}
        \caption{Similar as Fig.~\ref{pair-extra-1-8-ele} but for $1/8$ hole doping.} 
	\label{pair-extra-1-8-hol}
\end{figure*}

\begin{figure*}[t]
        \centering{
	\includegraphics[width=0.32\linewidth]{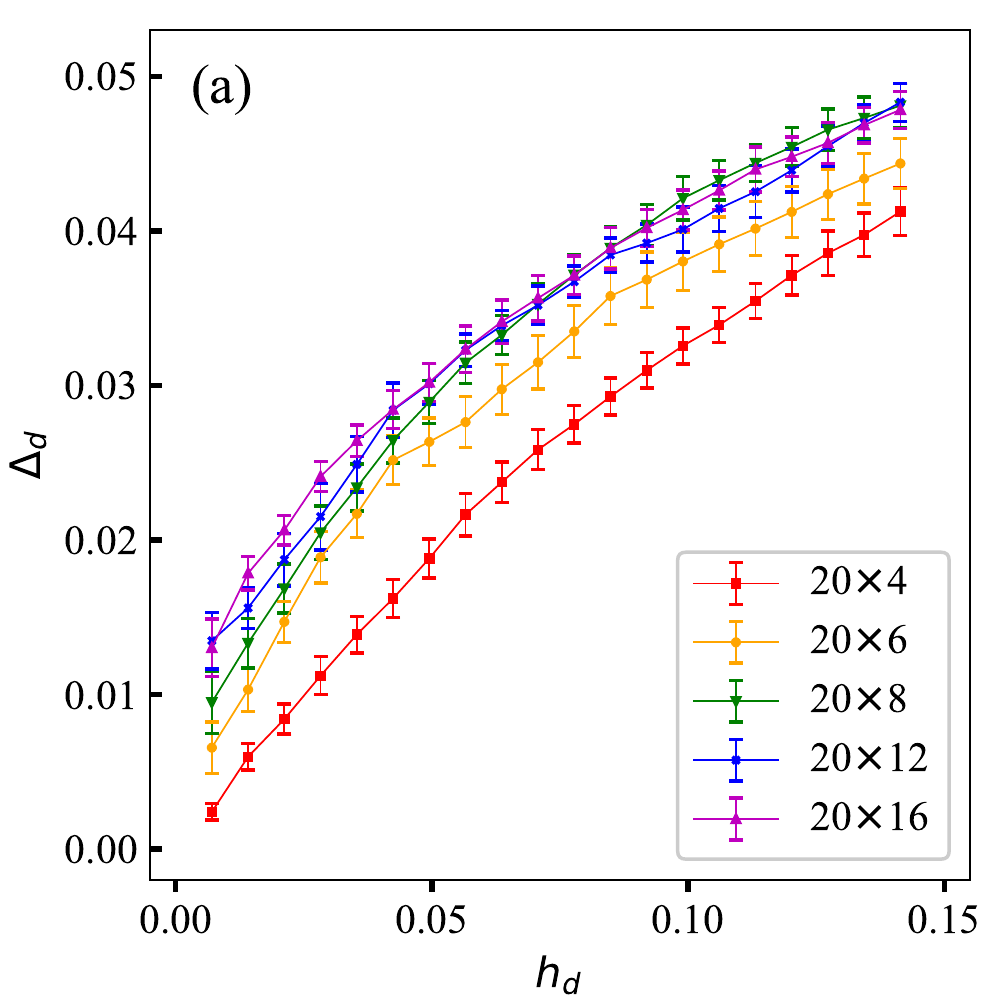}
        \includegraphics[width=0.32\linewidth]{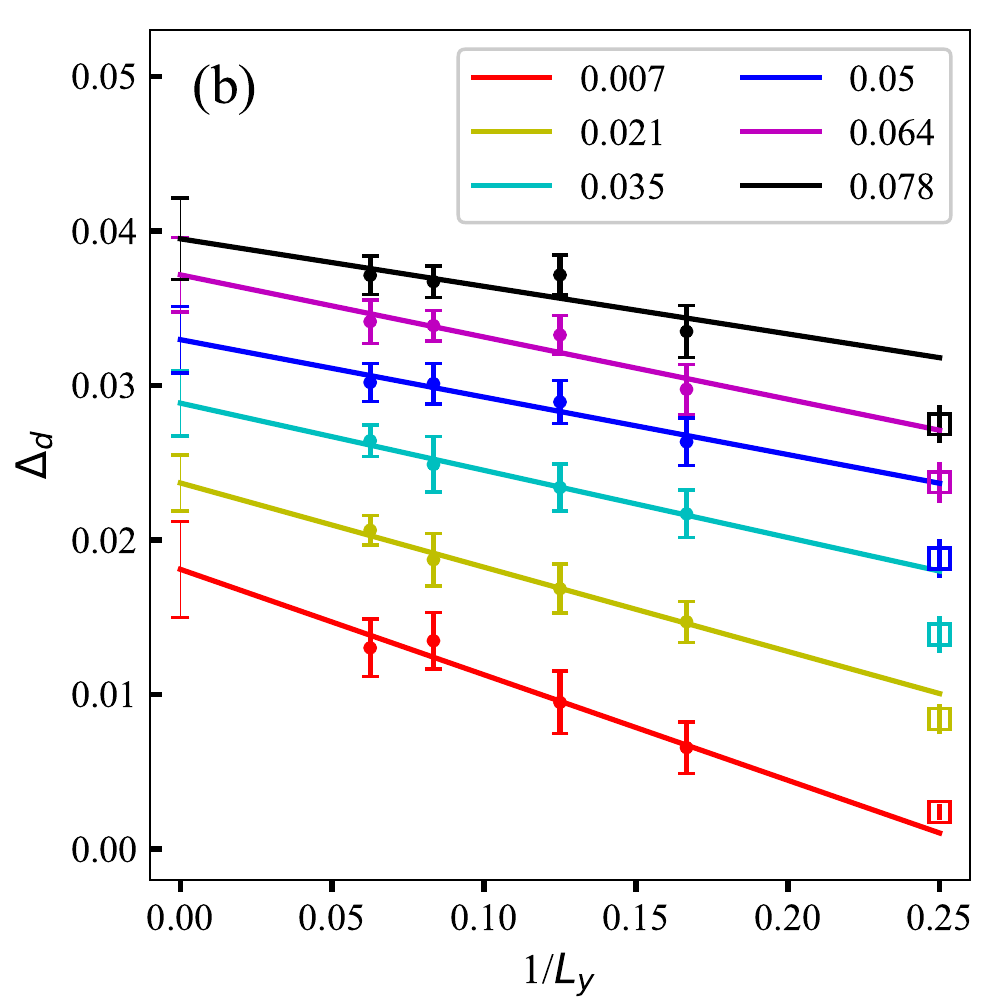}
        \includegraphics[width=0.32\linewidth]{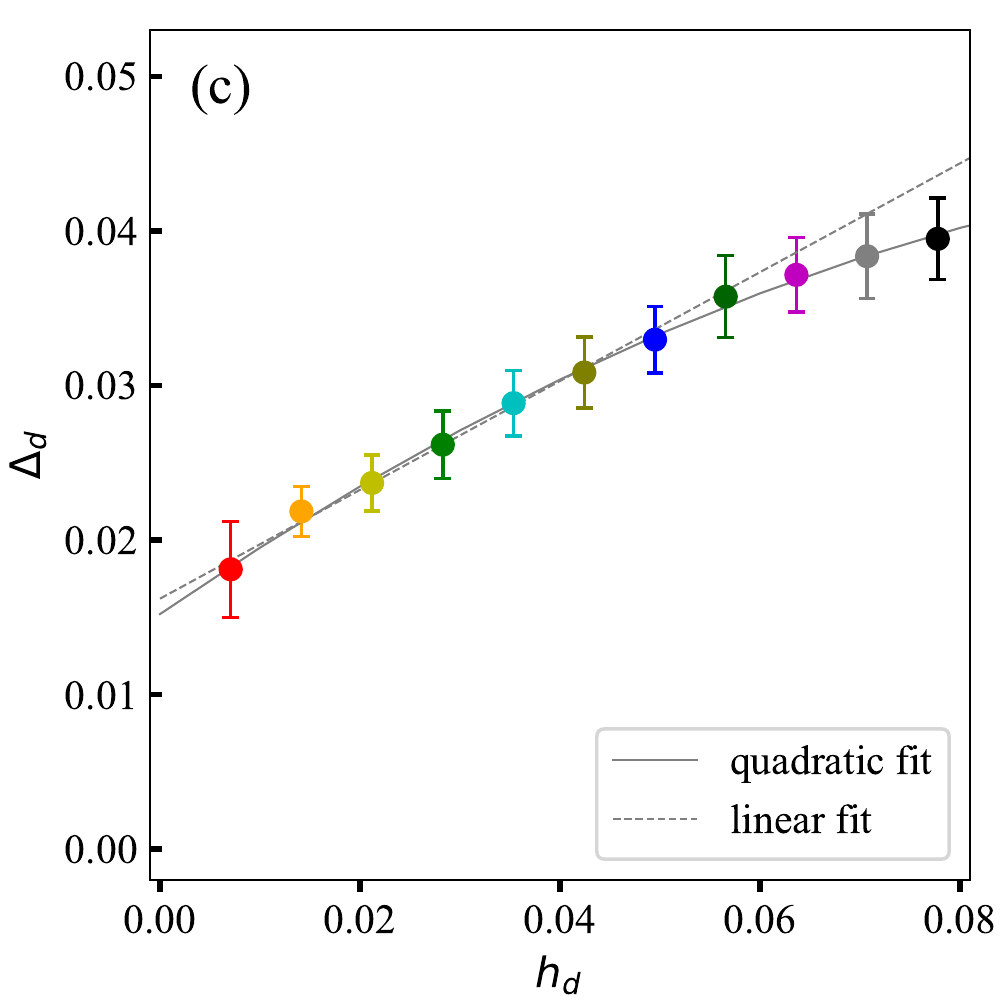}
	}
        \caption{Similar as Fig.~\ref{pair-extra-1-8-ele} but for $1/5$ hole doping.} 
	\label{pair-extra-1-5-hol}
\end{figure*}

\begin{figure*}[t]
        \centering{
	\includegraphics[width=0.32\linewidth]{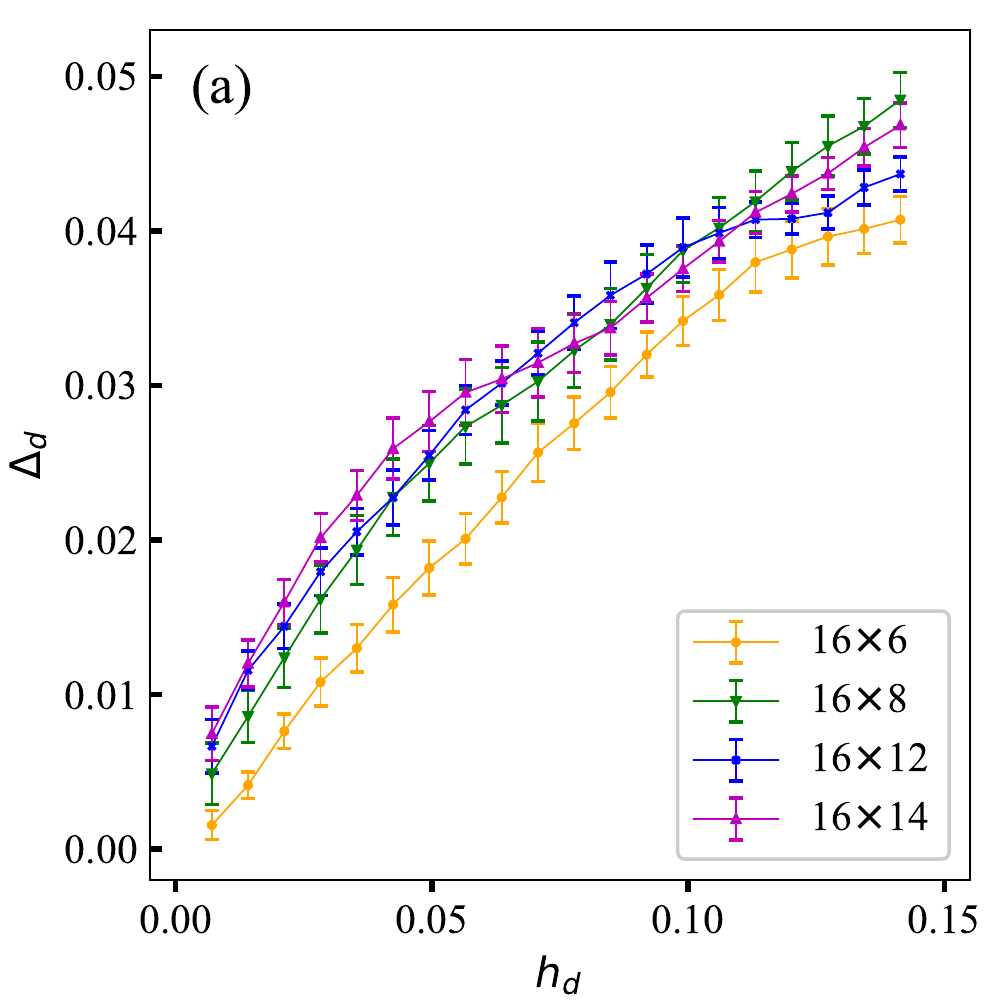}
        \includegraphics[width=0.32\linewidth]{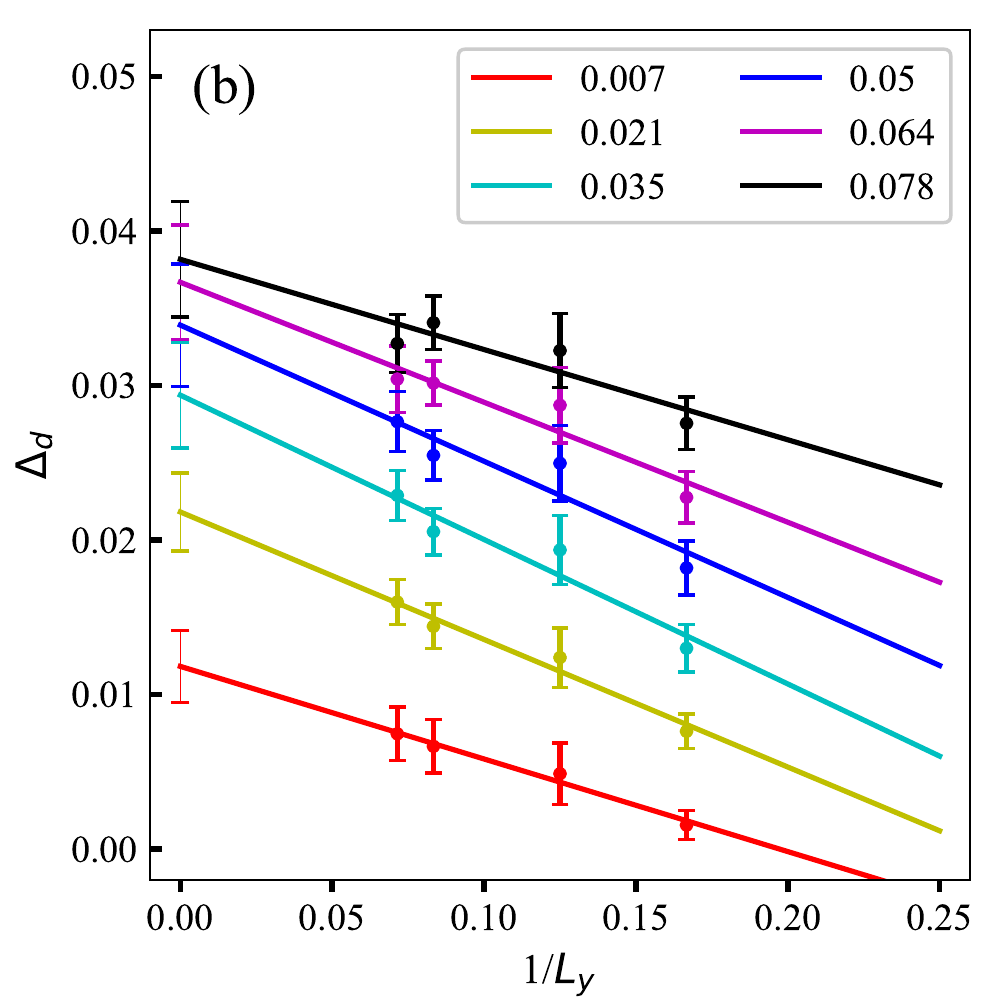}
        \includegraphics[width=0.32\linewidth]{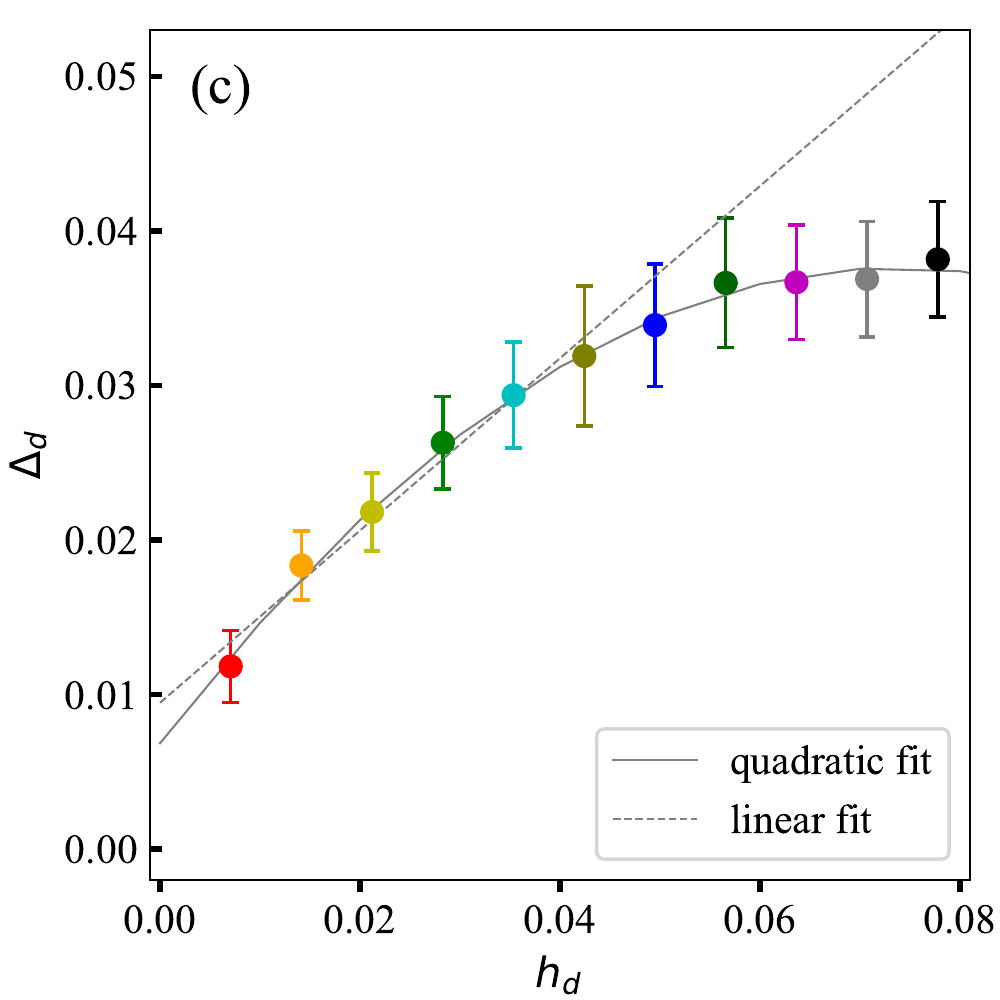}
	}
        \caption{Similar as Fig.~\ref{pair-extra-1-8-ele} but for $1/4$ hole doping.} 
	\label{pair-extra-1-4-hol}
\end{figure*}

\begin{figure*}[t]
        \centering{
	\includegraphics[width=0.32\linewidth]{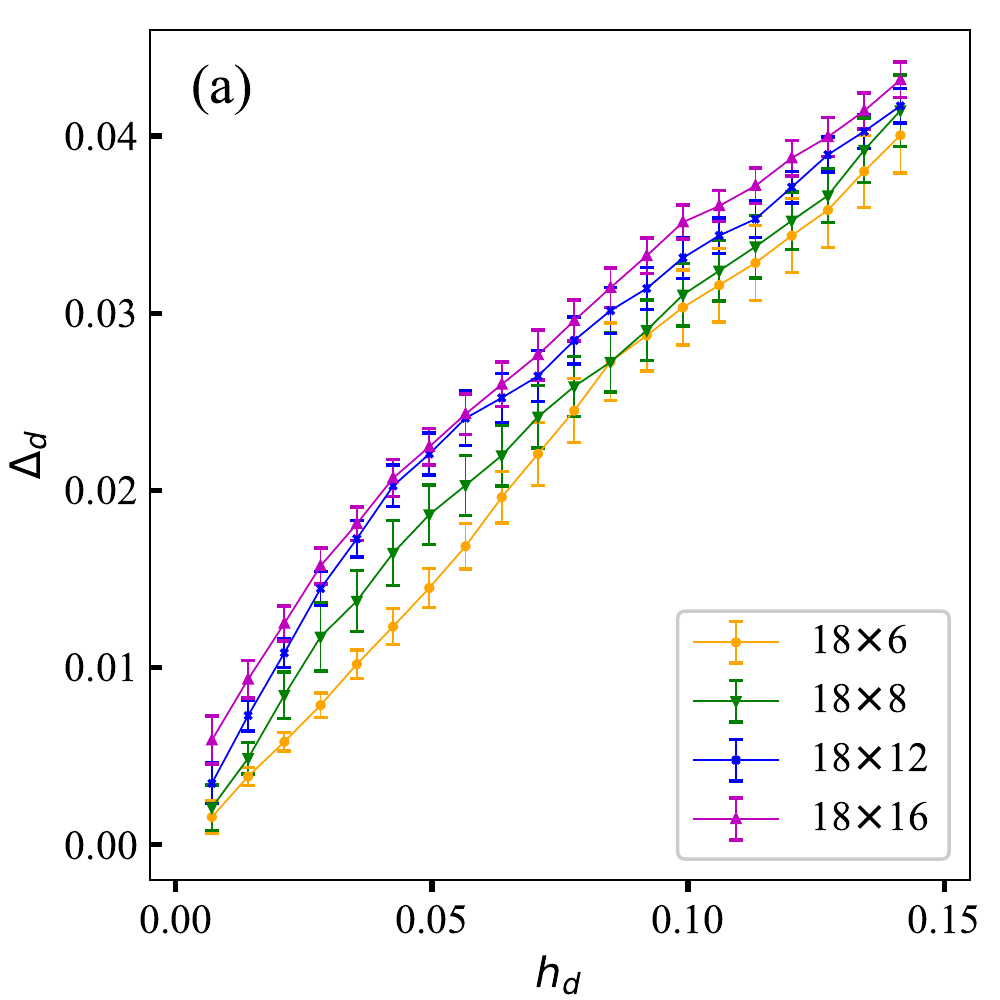}
        \includegraphics[width=0.32\linewidth]{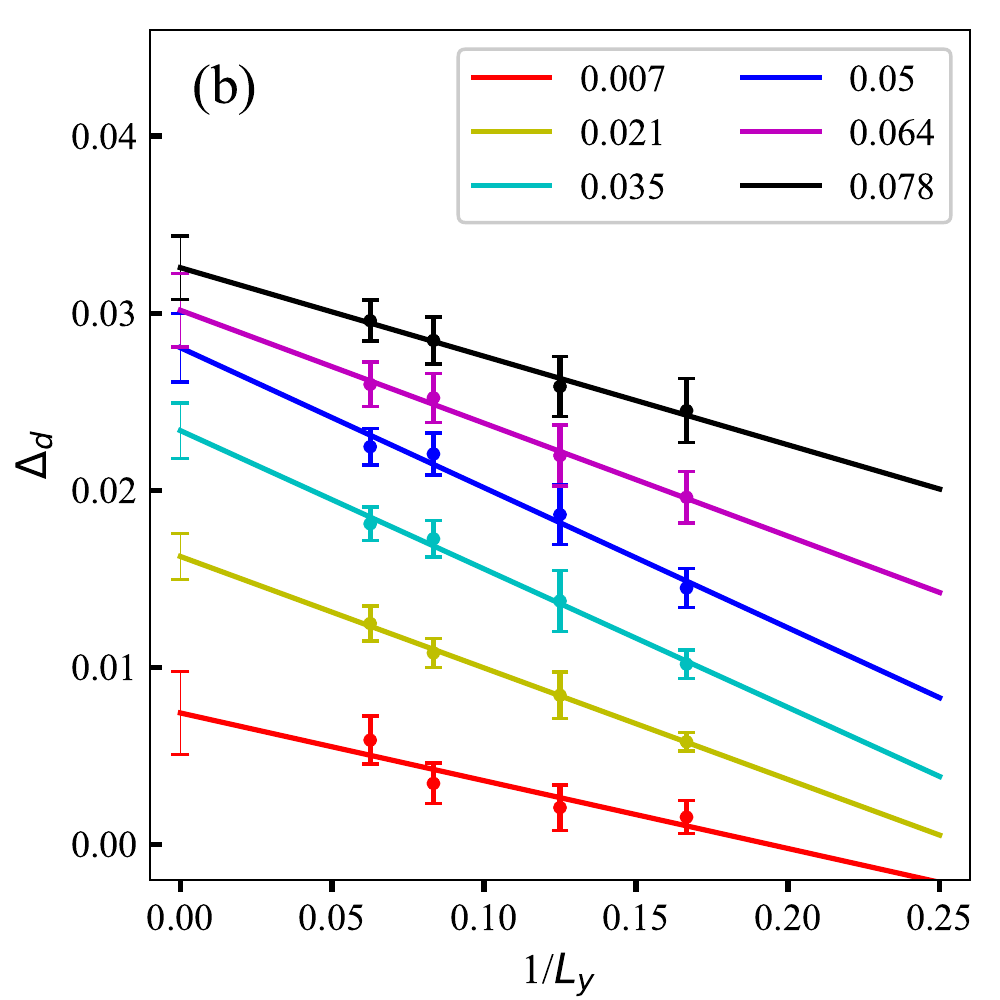}
        \includegraphics[width=0.32\linewidth]{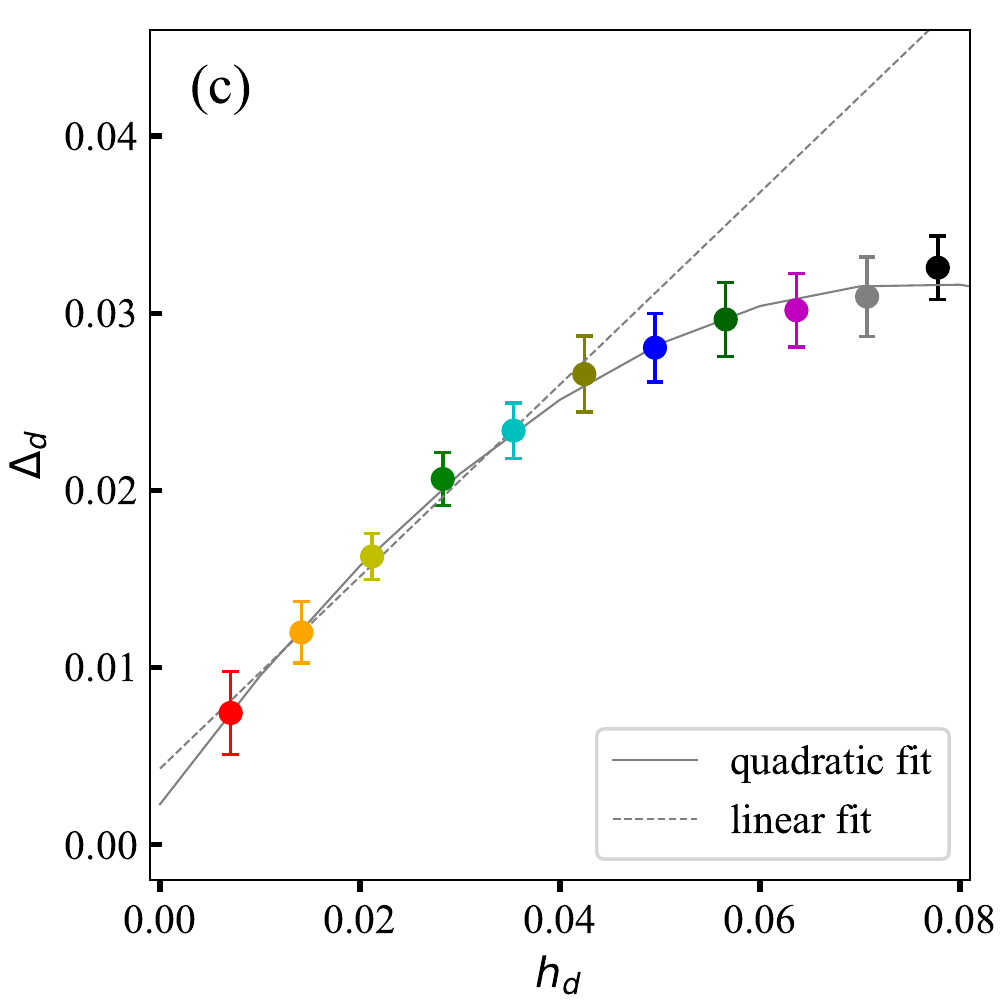}
	}
        \caption{Similar as Fig.~\ref{pair-extra-1-8-ele} but for $1/3$ hole doping.} 
	\label{pair-extra-1-3-hol}
\end{figure*}

In 
Figs.~\ref{pair-extra-1-8-hol}, \ref{pair-extra-1-5-hol}, \ref{pair-extra-1-4-hol}, and \ref{pair-extra-1-3-hol}
we present the data for 
pairing order in the hole-doped region, with $\delta = 1/8, 1/5, 1/4$, and $1/3$ respectively.
The final pairing order at TDL
for $\delta = 1/8, 1/5, 1/4$, and $1/3$ are $0.017(3),0.016(2),0.007(3)$, and $0.002(3)$, 
as shown in Fig.~1 in the main text.

\bibliography{supplement}